\documentclass{aa}  
\usepackage[utf8]{inputenc}
\usepackage{natbib}
\usepackage{hyperref}
\usepackage{threeparttable}
\usepackage{gensymb}
\usepackage{tabularx}
\usepackage{booktabs}
\usepackage{amsmath}
\usepackage{amssymb}
\usepackage{version}
\usepackage{graphicx}
\usepackage{pdfpages}
\usepackage{setspace}
\usepackage{booktabs}
\usepackage{multirow}
\usepackage{comment}
\usepackage{appendix}
\usepackage{txfonts}
\hypersetup{
    colorlinks=true,
    linkcolor=blue,
    filecolor=magenta,      
    urlcolor=cyan,
    citecolor=blue,
}
\begin{document} 

\title{Spectral analysis of the low-mass X-ray pulsar 4U 1822-371: Reflection component in a high-inclination system}
   
\author{A. Anitra \inst{1}, T. Di Salvo\inst{1}, R. Iaria\inst{1}, L. Burderi \inst{2} A.F. Gambino\inst{1}, S.M. Mazzola \inst{1}, A. Marino\inst{1}, A. Sanna \inst{2}, A. Riggio \inst{2}}

   \institute{ Dipartimento di Scienze Fisiche ed Astronomiche, Università di Palermo, via Archirafi 36, 90123 Palermo, Italy
   \and 
   Dipartimento di Fisica, Università degli Studi di Cagliari, SP Monserrato-Sestu, KM 0.7, 09042 Monserrato, Italy
}

% \abstract{}{}{}{}{} 
% 5 {} token are mandatory
 
  \abstract
%contex
{The X-ray source 4U 1822-371 is an eclipsing low-mass X-ray binary and X-ray pulsar, hosting a NS that shows periodic pulsations in the X-ray band with a period of 0.59 s. The inclination angle of the system is so high (80-85\degree) that in principle, it should be hard to observe %should be impossible observing
both the direct thermal emission of the central object and the reflection component of the spectrum because they are hidden by the outer edge of the accretion disc. Despite the number of studies carried out on this source, many aspects such as the geometry of the system, its luminosity, and its spectral features are still debated.}
%aim
{Assuming that the source accretes at the Eddington limit, the analysis performed in this paper aims to investigate the presence of a reflection component. No such component has been observed before in a high-inclination  accretion-disc corona source such as 4U 1822-371. To do this, we use large-area instruments with sensitivity in a broad energy range.} %never observed before in a  source like this.}
%Method
{We analysed non-simultaneous  \textit{XMM-Newton}  and  \textit{NuSTAR} observations  of  4U 1822-371 and  studied  the  average  broad-band spectrum. We first reproduced the results reported in the literature, then focused on the research of reflection features. We modelled the spectral emission of the source  using two different reflection models, \textsc{diskline} plus \textsc{pexriv} or, alternatively, the self-consistent reflection model \textsc{RfxConv}. We also included six Gaussian components ascribable to emission lines %We use a model that includes the self-consistent reflection model RfxConv to which we add six different gaussian lines to fit the emission features
at low energies. }
%%From which we derive constraints on the parameters characterising the emission region of the different spectral components.}
%Result
{In our analysis, we find significant evidence of a reflection component in the spectrum, in addition to two narrow (Gaussian) lines at 6.4 and 7.1 keV associated with neutral (or mildly ionised) iron, Fe K$_{\alpha}$, and K$_{\beta}$ transitions, respectively. The continuum spectrum is well fitted by a saturated Comptonisation model with an electron temperature of 4.9 keV and a thermal black-body-like component that might be emitted by the accretion disc at a lower temperature ($\sim 0.2$ keV). We identify  emission lines from O VIII, Ne IX, Mg XI, and Si XIV. We also added two new eclipse times related to \textit{NuSTAR} and Swift observations to the most recent ephemeris reported in literature, updating thus the ephemeris and finding a $P_{orb}$ = 5.57063023(34) hr and a $\dot{P}_{orb}$ value of 1.51(5) $\times$ 10$^{-10}$ s s$^{-1}$.}
%Conclusion
{In our proposed scenario, 4U 1822-371 is accreting at the Eddington limit with an intrinsic luminosity of $\sim 10^{38}$ erg/s, while the observed luminosity is two orders of magnitude lower because of the high inclination angle of the system. Despite this high inclination, we find that a reflection component is required to fit residuals at the Fe line range and to model the hard excess observed in the \textit{NuSTAR} spectrum. The inclination inferred from the reflection component is in agreement with values previously reported in literature for this source, while the best-fit value of the inner disc radius is still uncertain and model dependent. 
%%, $\sim 15\, R_g$, is significantly smaller than the proposed value for the magnetospheric radius (45 - 75 $R_{g}$) in this source. This would imply a higher accretion rate and/or lower magnetic field than what is assumed. 
More observations are therefore needed to confirm these results, which can give important information on the central emitting region in this enigmatic and peculiar source.
}
       %% LE KEYWORDS VANNO SCRITTE CORRETTAMENTE: COPIA DA ALTRI PAPER....
\titlerunning{Spectral analysis of the low-mass X-ray pulsar 4U 1822-371: Reflection component in a high inclination system}
   \authorrunning{ A.Anitra}

   \maketitle
%
%-------------------------------------------------------------------

\section{Introduction}
The source 4U 1822-371 is an eclipsing low-mass X-ray binary system (LMXB)  with an orbital period of 5.57h. It hosts a neutron star (NS) that shows periodic pulsations in the X-ray band (\citealt{Iaria}) with a period of 0.59 s. 

As estimated by \cite{Mason} using infrared observations, the inclination angle with respect to our line of sight lies in the range 80\degree - 85\degree. This high inclination angle implies that the direct emission from the innermost region should be shaded by the swelling in the external region of the accretion disc, caused by the inflowing matter transferred by the secondary star that impacts the outer disc. This source is considered the prototype of the accretion-disc corona (ADC) sources. This describes a particular system geometry in which most of the observed radiation emitted by the accretion flow is scattered by a corona that is formed by the evaporation of matter from the surface of the accretion disc, which leads to the accumulation of particles above and below the disc, forming a sandwich structure around it \citep{1973Shakura_Sunyaev}.
%%, therefore the only thermal component that can be observed, must come from the disc.\\
Another interesting feature of 4U 1822-371 is its activity as an X-ray pulsator. Analysing RXTE data, \cite{Jonker_2001} discovered coherent X-ray pulsations at $\sim 0.59$ s and inferred a pulse-period derivative of (-2.85 $\pm$ 0.04) $\times$ 10$^{-12}$ s/s, indicating a spin-up of the NS. Taking the observed luminosity of 10$^{36}$ erg/s estimated by \cite{Mason} for a distance of the source of 2.5 kpc into account, the authors also calculated the strength of the NS magnetic field. Using the relation of \cite{Ghosh_1979} that links the luminosity and the spin-up rate, they found an extremely high value of 8 $\times$ 10$^{16}$ G for the magnetic field, suggesting that the luminosity might be underestimated.%, that suggests the observed luminosity is probably underestimated.\\

\cite{Burderi_2010} studied the orbital evolution of the system and refined the orbital ephemeris of the source. They concluded that the orbital period increases at a rate of  $\dot{P}_{orb} = 1.50(7)$ $\times$ 10$^{-10}$ s/s \citep[see also][]{Parmar}. This high orbital period derivative cannot be explained by a conservative mass transfer at the accretion rate inferred from the observed source luminosity. According to these authors, a highly non-conservative mass transfer is required at a rate up to seven times the Eddington limit for an NS with a mass of 1.4 M$_{\odot}$ \citep[see e.g.][]{Burderi_2010, Bayless2010, Iaria2011, MAzzola}, indicating that a large part of the matter that is pulled off by the companion star is ejected from the system. The luminosity produced by this accretion process is about 10$^{38}$ erg/s, which differs from the luminosity reported by \cite{Mason}, but the magnetic field strength inferred with this luminosity implies a more reasonable value of 8 $\times$ 10$^{10}$ G \cite[see also][]{Jonker_2001}.
 
The spectral shape of the X-ray emission is still debated. %because of the high inclination angle, 
The main observed spectral component is a Comptonisation spectrum, which is the result of the inverse Compton scattering of soft thermal photons that are emitted by the compact object or by the accretion disc off the hot electrons that form the innermost part of the Corona around the NS.
\cite{Hellier_Mason_1989} fitted an EXOSAT spectrum of 4U 1822-371 using a power law and a black-body component (peaked at 1.8-2.0 keV) in addition to an iron line. They measured an apparent radius for the region of the black-body emission of about 0.25 km, which is much smaller than the NS radius. 

Recently, \cite{Iaria} performed an orbital phase-dependent spectral analysis of \textit{Chandra} and \textit{XMM-Newton} observations of 4U 1822-371, fitting the spectrum with a black-body at a temperature of 0.061 keV and with a Comptonised component characterised by an electron temperature of 3.01 keV and emitted by an optically thick (inner) corona having an optical depth of $\tau$ = 19.1. Both these components are partially absorbed by local neutral matter and interstellar medium (ISM). Several emission lines are observed at low energies,  probably emitted by plasma in the bulge at the outermost region of the disc, in addition to two narrow fluorescence iron lines associated with Fe XXV and Fe XXVI, which are produced in the innermost regions. %%that show no relativistic smearing.\\
The authors proposed that the partial covering component observed in the source spectra may be ascribed to local neutral matter around the system that cannot accrete onto the NS because it is emitting at the Eddington limit. They also ascribed the observed low luminosity of 10$^{36}$ erg/s to the high inclination of the source; an extended, optically thin corona ($\tau$ = 0.01) that scatters a small fraction of the inner emission along the line of sight may explain why we observe this low luminosity from the source. \\
In this paper we investigate the geometry of the system through a spectral analysis of X-ray data collected from non-simultaneous \textit{XMM-Newton} and \textit{NuSTAR} observations, taking advantage of the large effective area in a broad energy range and moderately good energy resolution at the iron-line energy. We find that the iron line may indeed be a blending of lines from iron in different ionisation states. Moreover, the large width of one of these line components and the Compton hump at high energies suggest  reflection, which might represent the first evidence of a reflection component for a high-inclination system such as 4U 1822-371.

\section{Observations}
The source 4U 1822-371 was observed by the \textit{XMM-Newton} observatory on 2017 March 3 between 01:10:54 UTC and 19:12:27 UTC (ObsId. 0784820101) for a duration of 69  ks.
During this observation, the EPIC-pn and MOS2 cameras were operated in timing mode with a medium filter, while the  MOS1 camera was set in small-window mode with a thick filter. The timing mode is useful to avoid photon pile-up: this effect occurs when the observed source is bright enough for different X-ray photons to arrive in a single pixel within a frame time, causing the instrument to register them as a single photon with an energy equal to the sum of the energies of the original photons. This causes a loss in photon counts as well as a hardening of the spectrum.
We verified that MOS 1 was affected by pile-up because the camera was operated in imaging mode during the observation, and we found that the data collected by the Epic-pn and the MOS 2 were incompatible below 3 keV due to detector instrumental features. For these reasons, and because Epic-pn has a greater effective area than the MOS cameras in the same energy range, we preferred to use the Epic-pn data for our analysis. To obtain a better coverage of the spectrum at the lower energies, we also used  data collected by the two RGS cameras, which provide a high-energy resolution in the range between 0.35 keV and 2.5 keV.

We reduced the \textit{XMM-Newton} data using the Science Analysis Software (SAS) v17.0.0. We first extracted the images from the event files to select the region from which events are collected. For the EPIC-pn data, we selected the columns in the interval $26 <RAWX<47$ for the source and $3<RAWX<5$ for the background, respectively. For the RGS background, we selected a region in CCD9 in which generally the background dominates because it is located close to the optical axis and records fewer source photons. 
We accumulated the EPIC-pn 0.3-10 keV light curve with a bin time of 100 seconds considering $PATTERN <=4$ to extract only single and double good events, and $FLAG=0$ to avoid events from the pixels at the edges of the CCD. In the light curve of the source, we found three partial eclipses separated by a period of about 20 ks; in order to exclude the time intervals during which the luminosity is reduced from the average spectrum, we created a set of good time intervals (GTI) to exclude the eclipses from the following spectral analysis. We eventually extracted source and background spectra and combined the RGS1 and RGS2 spectra, first and second order.

The \textit{NuSTAR} observation was performed between 2018 April 25, 14:26:09, and April 27, 07:10:26 (ObsID 30301009002), for a total exposure time of 29 ks. We used the \textsc{Nupipeline} and \textsc{Nuproducts} scripts within HEASOFT to obtain the average spectra: for both the source and background of FPMA and FPMB, we selected a circular region with a radius of 30 arcsec. As done for EPIC-pn data, we produced light curves and created the GTI files, after which we extracted the source and background (out-of-eclipse) average spectra, the ancillary files, and the response matrices for each of the two \textit{NuSTAR} telescopes. We rebinned these spectra in order to have a minimum of 30 counts per bin.
The \textit{NuSTAR} spectra provide a better coverage of the spectrum at high energies: %thanks to its wide energy range (3-80 keV):
this feature makes \textit{NuSTAR} a perfect complement for our work because together with the low-energy coverage provided by \textit{XMM-Newton}, it allows a detailed analysis of all the main features (discrete features as well as the continuum) that characterise a reflection spectrum emitted by X-ray binaries.

\section{Timing analysis}
\begin{figure}  
    \centering
    \includegraphics[width=.5\textwidth]{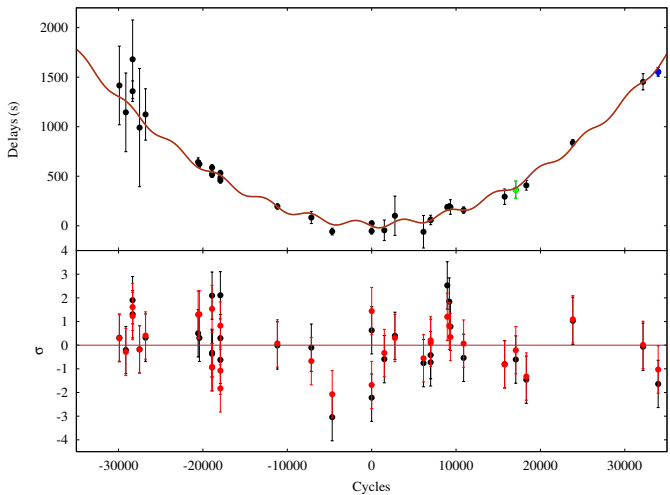}
        \caption{\label{timefigure} Delays of the eclipse arrival times as a function of the orbital cycle for the LQS model (top panel). Residuals in units of $\sigma$, obtained with the cubic (black) and LQS (red) model, respectively (bottom panel).}
\end{figure}

\begin{table}[]\centering 
\caption{\label{timetable}Top part: Eclipse arrival times. Bottom part: Best-fit values for the parameters derived with the different models described in the text (quadratic, cubic, and LQS).}
\resizebox{0.8\textwidth}{!}{\begin{minipage}{\textwidth}
 
\begin{tabular}{lccc}
\toprule
Eclipse time  & Cycles & Delays (s) & Satellite\\
\toprule
54317.0587(10)  & 17078 & 364(87) & Swift \\
58234.1536(5) & 33954 & 1554(43) & NuSTAR \\
\\
\toprule
 Parameter & Quadratic & Cubic & LQS  \\ 
\toprule 
a (s) &         12   $\pm$ 12 & -3    $\pm$ 12  & 13   $\pm$ 11\\
b (10$^{-4}$ s) &     -11   $\pm$ 5 & 19   $\pm$ 12   & -5   $\pm$ 6  \\
c (10$^{-6}$ s) &         1.42   $\pm$ 0.04 & 1.51 $\pm$ 0.05   & 1.41 $\pm$ 0.04 \\ \\
d (10$^{-12}$ s) &        -        &   -6    $\pm$ 2   & - \\

A (s) &         -    & - & 34   $\pm$ 11    \\
N$_{Mod}$ ($10^{-3}$) &         -    & - & 5.4 $\pm$ 0.1 \\
N0 ($10^{-3}$) &         -    & - & -3.1  $\pm$ 0.3 \\
P$_{Mod}$ (years) &         -    & - &  3.43 $\pm$ 0.07 \\ \\

T$_{0}$ (TJD) &         50353.08741(14)  & 50353.08725(14) & 50353.08744(13) \\
P$_{0}$ (days) &         0.232109559(6)  & 0.232109593(14) & 0.232109565(7)   \\
$\dot{P}$ (10$^{-10}$ s s$^{-1}$) &         1.42(3)  & 1.51(5) & 1.41(4) \\
$\ddot{P}$ (10$^{-19}$s s$^{-2}$) &         - & -0.94(35)  & - \\ \\
\hline 
$\chi^{2}$(dof) & 47.43(31)&   38.21(30)   &   34.30(28) \\
\hline
\end{tabular}\end{minipage}}
\end{table}

The \textit{NuSTAR} ObsID 30301009002 represents the most recent observation of 4U 1822-371, and as explained in the previous section, the extracted light curve shows an eclipse that has never been used to perform an analysis aimed at studying the orbital parameters of the system.
For this reason, we performed a timing analysis in order to investigate whether it is possible to extend the most recent orbital ephemeris of the system reported by \cite{MAzzola} and improve the accuracy of the orbital parameters.

We applied the barycentric correction to the event file, extracted as discussed in section 2, using the Ftool \textsc{barycorr}. We folded the light curve obtained from the two \textit{NuSTAR} cameras, FPMA and FPMB, using as reference time $T_{fold}$ = 50353.08728 MJD  and a trial orbital period of P$_{fold}$=0.232109571 days \citep{MAzzola}. The eclipse arrival time was evaluated by using the method described by  \cite{Burderi_2010}, finding a value of T$_{ecl}$ = 58234.1536(5) MJD/TDB. 

We extended the ephemeris of the source with a further eclipse arrival time extracted from a \textit{Swift} observation performed between 2007 August 3, 01:20:01, and  2007 August 5, 21:07:22 (ObsID 00036691002 and 00036691003). Using the same method as adopted for the \textit{NuSTAR} observation, we find an arrival time of T$_{ecl}$ = 54317.0587(10) MJD/TDB.  

%We then calculated the delays and the orbital cycles related to the two new eclipse times: we subtracted from our measures the reference eclipse time T$_{0}$ = 50353.08728 MJD and divided the result from the reference orbital period of P$_{0}$ = 0.232109571 s; the integer part N of the number obtained is the number of orbital cycles since T$_{0}$, while the decimal part, multiplied by the reference period P$_{0}$, is the delay.
The inferred delays of the eclipse arrival times with respect to the adopted ephemeris and the corresponding number of orbital cycles are reported in Table \ref{timetable}.
We fitted the delays as a function of the cycles using three different models. We first modelled the data with the quadratic function $y= a + bN + cN^{2}$, where the parameters \textit{a} and \textit{b} represent the correction for the value of the eclipse time ($\Delta T_{ecl}$) and the orbital period ($\Delta P/P_{0}$), respectively, while the term \textit{c} describes the variation in orbital period and allows us to determine its derivative ($P_{0}/2\dot{P}$). We obtained a $\chi^{2}$ ($d.o.f.$) = 47.43(31) and  $\dot{P}$ = 1.42(3) $\times 10^{-10}$ s/s. The  results are reported in Table \ref{timetable}.

To improve the fit, we added a cubic term $(y= a + bN + cN^{2} + d N^{3})  $ to the previous model, where the new coefficient describes the orbital period second derivative ($P^{2}_{0} \ddot{P}_{orb}/6$). The model improves the fit and achieves a $\chi^{2}$ ($d.o.f.$) = 38.21(30). This allows us to infer   $\dot{P}$ = 1.51(5) $\times$ 10$^{-10}$ s/s and $\ddot{P}$ = -0.94(35) $\times$ 10$^{-19}$ s/$s^{2}$ , in line with previous results \citep{MAzzola}. We tested the improvement of the fit with an \textsc{F-test} that compared the quadratic and cubic models, and obtained a probability of chance improvement of about 0.01. This indicates that the addition of the cubic term is significant at 99 $\%$ confidence level. Adopting the cubic ephemeris, we updated the orbital ephemeris of the source as
\begin{align*}
            T_{ecl} =  50353.08725(14) MJD/TDB + 0.232109593(14)N \\
             + 1.752(29) \times 10^{-11} N^{2} -8.44(52) \times 10^{-22} N^{3} \;.
\end{align*}
\cite{MAzzola} inferred a possible sinusoidal modulation of the delays. To investigate this, we used a model composed of a quadratic and sinusoidal function, $LQS= a + bN + c N^{2} + A sin[2\pi (N-N_{0})/N_{mod}]$, where A is the semi-amplitude of the modulation. The best-fit parameters are reported in the third column of Table \ref{timetable}, and in Figure \ref{timefigure}, the residuals of the cubic and LQS model are compared. We obtained a $\chi^{2}$ ($d.o.f.$) =  34.30(28) and an \textsc{F-test} probability of chance improvement of about 0.03 with respect to the quadratic model, meaning that the sinusoidal modulation improves the fit at the 97\% confidence level. The origin of this modulation is still debated and may be associated with the gravitational quadrupole coupling (GQC) mechanism through tidal interaction or with a third body in the system \citep{MAzzola}. Future observations may allow us to further extend the ephemeris, confirming or disproving this modulation, and possibly explaining its origin.
\section{Spectral analysis}
We initially fitted the data of the two non-simultaneous observations separately in order to understand their spectral features in detail and determine whether the two datasets were compatible with each other.
We decided to adopt the energy range 4 - 45 keV for the \textit{NuSTAR} spectrum and 0.6 - 10 keV for the \textit{XMM-Newton} spectrum. We dropped the \textit{NuSTAR} data above 45 keV and below 4 keV because at higher energies, the background spectrum overcomes the source, and at lower energies, the spectrum diverges due to a flux excess in the FPMA camera \citep{Madsen2020}. For \textit{XMM-Newton,} we considered the RGS data only between 0.6 - 2 keV because above this threshold, the instrument is not well calibrated, and below it, some systematic features are expected, for instance, at the oxygen edge. In addition, we ignored EPIC-pn data lower than 2 keV because of some systematic features at 1.8 keV (caused by an instrumental Si edge).

\subsection{Comparison with a previous spectral decomposition}
We initially used the model proposed by \cite{Iaria}, which presents the most recent spectral analysis of 4U 1822-371 in the literature to date, with the aim to test whether a model without a reflection component can adequately describe our spectra. The model comprises a Comptonised component (\textsc{CompTT} in XSPEC) that is partly absorbed by neutral local matter and partly by the ISM. In this case, we assumed the same abundance (\citealt{Asplund_2009}) and the same cross-section (\citealt{Verner_1996}) as were chosen \cite{Iaria}, with the aim to compare our results to those available in the literature. For both datasets, we took possible inter-calibration issues into account by including a constant multiplicative component, fixed to 1 for Epic-pn and for the \textit{NuSTAR} FPMA camera. The component was kept free to vary for the other instruments (note that the \textit{RGS} parameters are linked to those of \textit{pn,}  while the \textit{FPMB} parameters are linked to those of \textit{FPMA} ).
The contribution of the interstellar photoelectric absorption was modelled by the  \textsc{phabs} component, while to take the partial photoelectric absorption due to local matter into account, we used the model \textit{phabs$\ast$[c$\ast$cabs$\ast$phabs$\ast$(CompTT)+(1-c)$\ast$CompTT]}. 
The first term in the brackets represents the partially absorbed Comptonisation spectrum, where \textsc{c} is a constant limited between 0 and 1 that expresses the partial covering fraction. The second \textsc{phabs} component is the photo-absorption due to local matter, while \textsc{cabs}, which takes into account the effect of Thomson scattering, is linked to the second \textsc{phabs}, under the assumption that local matter mainly causes this effect.

To obtain a constraint on photoelectric absorption, we firstly fit EPIC-pn and RGS data together, taking advantage of the good spectral coverage of the RGS at low energies, where the effect of photoelectric absorption is more relevant. We used the same strategy to constrain the partial photo-absorption by the local matter in the system.
Because 4U 1822-371 has always shown a steady behaviour with little spectral variations, we do not expect a great difference between the parameters of the model of the two observations, although they were taken one year apart.
After a first attempt, we realised that several emission features were present in the \textit{XMM-Newton} and \textit{NuSTAR} spectra, so that we added three Gaussian lines at energies of 6.49 keV, 6.67 keV, and 7.1 keV to improve
the fit that are attributable to Fe I $K_{\alpha}$, Fe XXV, and Fe I $K_{\beta}$ emission, respectively.
In the \textit{NuSTAR} spectrum, the K$_{\alpha}$ emission lines are less distinguishable, and the energy of the line at 7.1 keV was fixed during the fit because it showed a tendency to overlap with the other Gaussian profiles.
 We also improved the model by adding four emission lines at lower energies to fit localised features in the \textit{XMM-Newton} band at 0.65 keV, 0.916, keV 1.346 keV, and 2.002 keV, , respectively. The centroids of the Gaussian components are in line with those reported by \citet{Iaria}, and we associate them with transitions of O VII, Ne IX, Mg XI, and Si XIV ions, respectively.

\begin{figure}
    \centering
    \includegraphics[width=.5\textwidth]{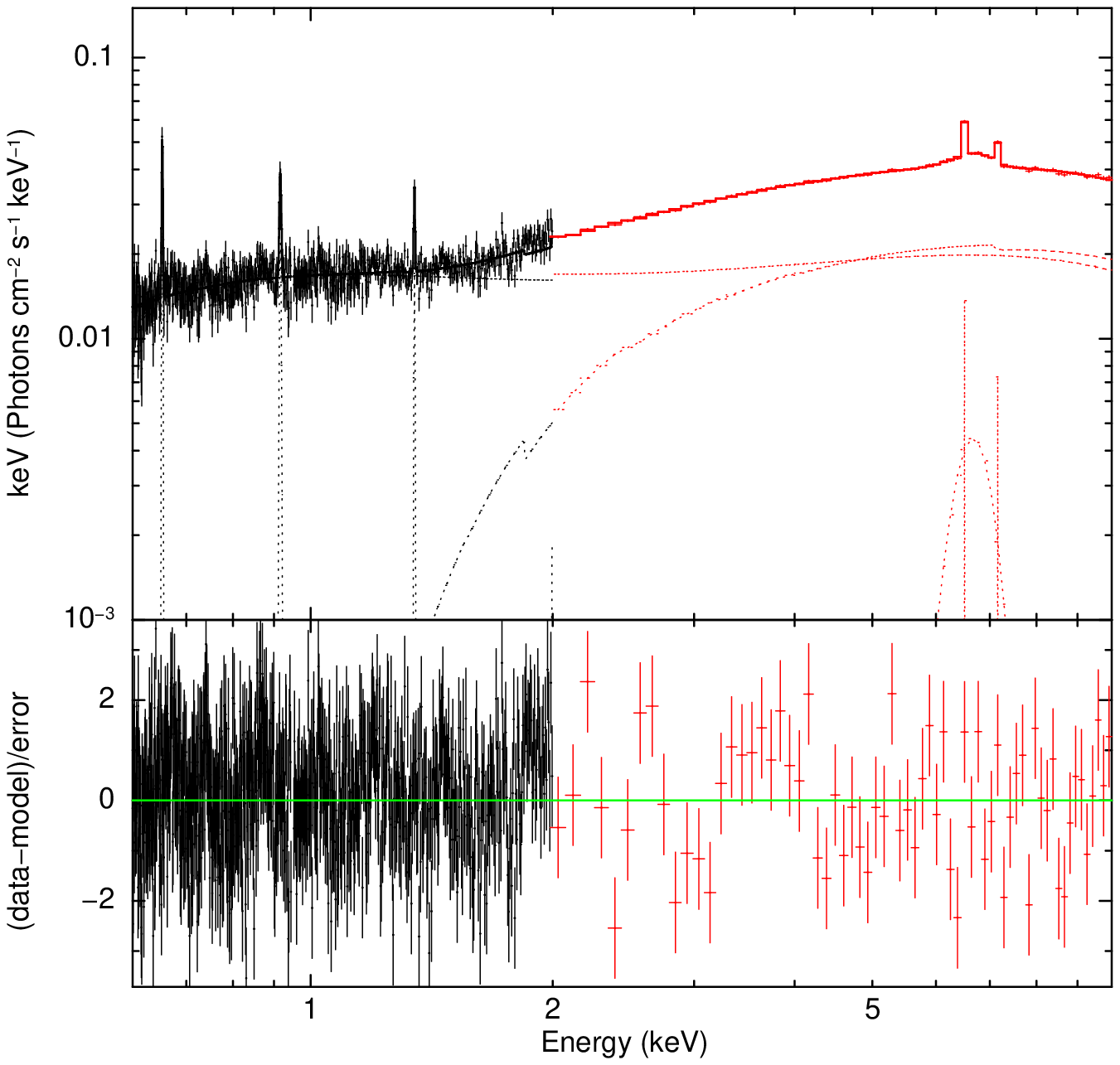}
\includegraphics[width=.5\textwidth]{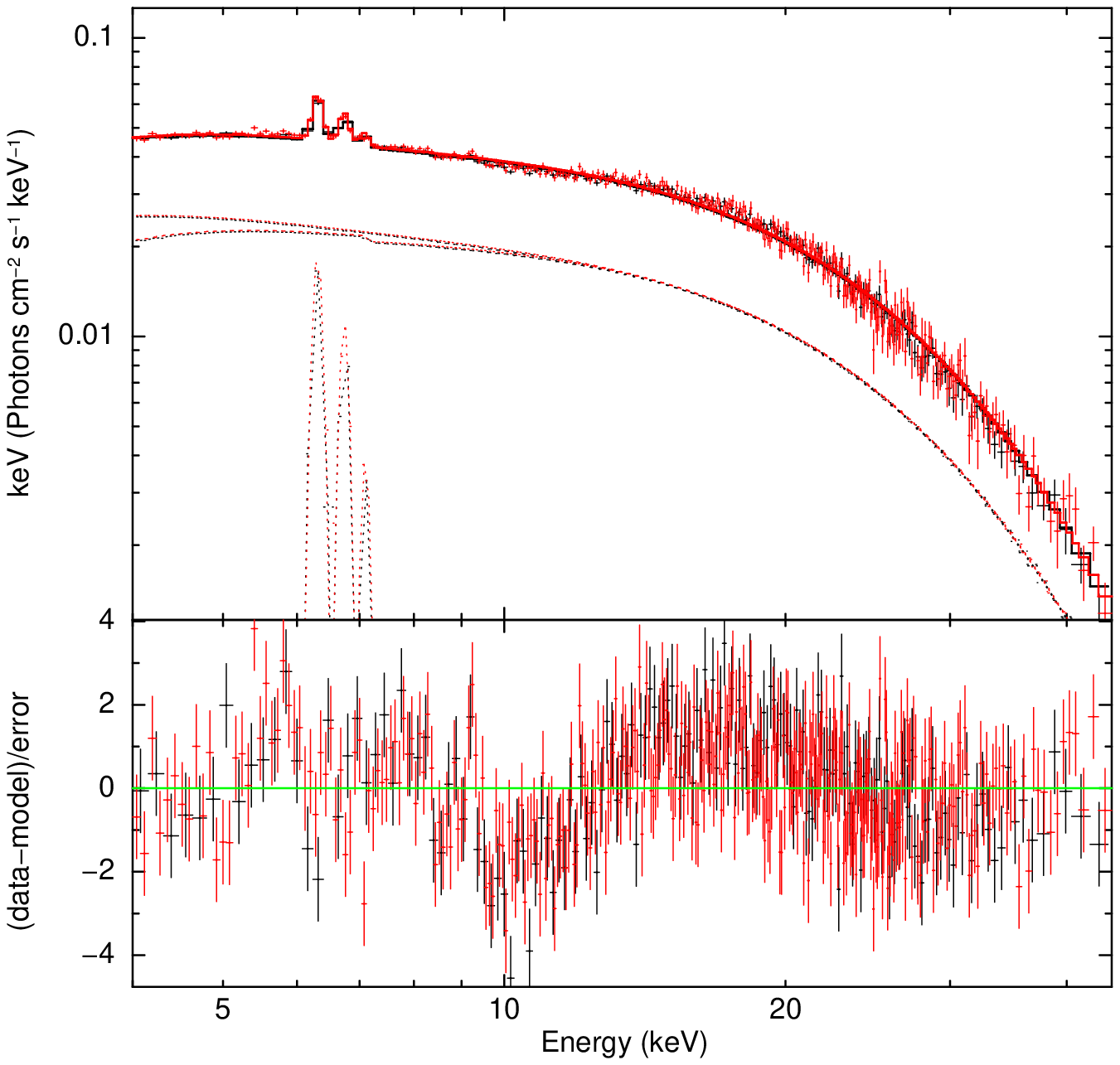}
    \caption{Spectra and residuals in units of sigma with respect to the model: $constant \cdot phabs [c\cdot cabs \cdot phabs \cdot (CompTT)+(1-c)CompTT + 7\;gaussian\;lines]$ for the \textit{XMM-Newton} spectrum (top panel) and \textit{NuSTAR} spectrum (bottom panel). Data were rebinned for visual purposes.}
    \label{fig:Iaria}
\end{figure}

The best-fit values obtained for the model parameters are shown and compared to the parameter reported by \cite{Iaria} in table \ref{Iaria_compared}. Only part of the values found for the \textit{XMM-Newton} spectra are consistent within the errors wit those reported previously. In particular, the equivalent hydrogen column density of the interstellar photoelectric absorption and the seed-photon temperature deviate from previous results.
From the fit of the \textit{NuSTAR} data, most of the parameters slightly deviate from those found in \citet{Iaria}. In addition to the parameters related to the components \textsc{phabs} and \textsc{cabs} (linked to the previous fit), the temperature of the electron cloud and seed photons reaches significantly higher values than what was found by \cite{Iaria} and for the \textit{XMM-Newton} spectrum. 
The best-fit normalisation value of the Gaussian at 7.1 keV, which is 18$\%$ of the value found for the Fe I emission line, as well as its centroid energy, are in agreement with the association of this line with the Fe I $K_{\beta}$ transition. 
Fig. \ref{fig:Iaria} shows that the \textit{NuSTAR} spectrum shows peculiar residuals in the iron emission region and from 10 to 40 keV. It is possible that the mismatch between the model and the data indicates that the spectrum needs a reflection continuum component because the latter residuals can be attributable to the Compton reflection hump, which is a typical signature of a reflection component.
\subsection{Testing a reflection component \label{reflection}}
To test the possible presence of a reflection component, we tried two different models: in \textsc{Model 1}, we used the self-consistent model \textsc{Rfxconv}, while in \textsc{Model 2,} the reflection is described by the combination of \textsc{diskline} and \textsc{pexriv} models.
%Added by TeX Support
\begin{table}[]\centering
 \caption{\label{Iaria_compared}Comparison of best-fit values for the parameters of the model described in the text, with those found by \cite{Iaria}.}
 \small\addtolength{\tabcolsep}{-5pt}

 \begin{threeparttable}
\begin{tabular}{llccc}
\toprule
\toprule
Parameter&  XMM\tnote{*} & NuSTAR\tnote{*} &  I-2013  \\
\toprule
\toprule

{\sc N$_{H_{phabs}}$} (10$^{22}$ cm$^{-2}$) &$0.010^{+0.003}_{-0.003}$ &0.01\tnote{$\dagger$} &  0.163$^{+0.008}_{-0.006}$\\
{\sc N$_{H_{pcf}}$} (10$^{22}$ cm$^{-2}$)  & $4.02^{+0.05}_{-0.08}$  &   4.02\tnote{$\dagger$}& 4.97 $\pm$ 0.12  \\
{\sc c} &      $0.538^{+0.002}_{-0.003}$ &   0.54\tnote{$\dagger$}               & 0.607 $\pm$ 0.009         \\

kT$_{0}$ (keV)  & $0.199^{+0.002}_{-0.003}$ &$ 0.85  \pm 0.02$   & 0.061 $\pm$ 0.003\\
kT$_{e}$ (keV)  &   $3.089^{+0.008}_{-0.013}$  & $5.05 \pm 0.03$        & 3.05 $\pm$ 0.04           \\
$\tau$   &   $22.92^{+0.04}_{-0.06}$ & 12.9$\pm$ 0.1    &21.0 $\pm$ 0.4        \\
N$_{Cp}$ (10$^{-2}$)        & 5.00$^{+0.007}_{-0.008}$  & 2.91 $\pm$ 0.04    &7.7 $\pm$ 0.2        \\

\hline \\
E$_{Line}$ (keV) & $ 6.48 \pm 0.01$&$6.31\pm 0.03$  & 6.402 $\pm$ 0.004 \\
$\sigma$ (keV) & $< 0.006$& 0.075\tnote{$\dagger$}  & 0.075 $\pm$ 8\\
N (10$^{-4}$) & $2.6^{+0.2}_{-0.1}$ & $ 5.9 \pm 0.5$ &  3.24 $\pm$ 0.13 \\
\\
E$_{Line}$(keV) & $6.65 \pm 0.03$& $6.74 \pm 0.05$  & 6.703 $\pm$ 0.013 \\
$\sigma$ (keV) & $0.37^{+0.02}_{-0.03}$& 0.0755\tnote{$\dagger$}  & 0.025\tnote{$\dagger$}\\
N (10$^{-4}$) & $5.9^{+0.3}_{-0.4}$& $3.3 \pm 0.5$ & 0.70 $\pm$ 0.08\\
\\
E$_{Line}$ (keV) & $7.10^{+0.02}_{-0.01}$ & 7.1  & 6.991 $\pm$ 0.009\\
$\sigma$ (keV) & $< 0.05$ & 0.075\tnote{$\dagger$} & 0.056$^{+0.008}_{-0.016}$\\
N (10$^{-4}$) & $ 1.3 \pm 0.2$&1.1 $_{-0.6}^{+0.4}$ & 1.38$^{+0.11}_{-0.06}$\\ \\
\hline
E$_{Line}$& $0.6535^{+0.0003}_{-0.0002}$ &-& 0.651794$^{+0.000336}_{-0.000004}$ \\
  $\sigma$ (keV)  & $0.0009^{+0.0004}_{-0.0003}$&-& < 0.0007 \\
  N(10$^{-4}$) & $1.6 \pm 0.2$ &-& 3.6 $\pm$ 0.4\\ \\

 E$_{Line}$ & $0.916  \pm 0.001$&-& 0.914 $\pm 0.001$\\
  $\sigma$ (keV)  & $0.002^{+0.0008}_{-0.0011}$ &-&0.004 $\pm$ 0.001\\
 N(10$^{-4}$) & $1.4^{+0.3}_{-0.2}$&-& 3.8 $\pm$ 0.3\\ \\ 
 
 E$_{Line}$& $1.346^{+0.003}_{-0.002}$ &-&1.343\tnote{$\dagger$}\\
 $\sigma$ (keV)  & $< 0.004$ &-& 0.003\tnote{$\dagger$}\\
  N(10$^{-4}$) & $0.7 \pm 0.2$ &-&0.6 $\pm 0.2$\\ \\

E$_{Line}$ & 2.002\tnote{$\dagger$}  &-& 1.984$^{+0.01}_{-0.006}$\\
$\sigma$ (keV)  & 0.004\tnote{$\dagger$} &-&0.004\tnote{$\dagger$}\\
 N(10$^{-4}$)  & $<0.4$ &-&1.5 $\pm 0.15$\\ \\

 \hline

$\chi_{red}^{2}(dof) $& 1.16(1544) & 1.31(1379) &  1.06(2400)\\
\hline
\end{tabular}
\begin{tablenotes}
\footnotesize
\item[$\dagger$]We decide to keep frozen these parameters to the values found for the XMM-newton spectrum.
\item[*] Model: \textsc{constant*phabs(c*cabs*phabs*(CompTT) + (1-c)CompTT+3gaussian)}.
\end{tablenotes}
\end{threeparttable}
\end{table} 
We initially modified the previous model by replacing \textsc{Comptt} with another Comptonisation model (\textsc{NthComp}), substituting the \textsc{phabs} model with \textsc{Tbabs}, that is, the Tuebingen-Boulder ISM absorption model, setting the most recent ISM abundance and cross-section \citep[see][]{2000ApJ...542..914W, Verner_1996}.
We decided to analyse the \textit{XMM-Newton} and \textit{NuSTAR} spectra simulatenously because the \textit{Epic-pn} and \textit{NuSTAR} data overlap in the iron emission-line region, which guarantees a better statistics. Moreover, by merging the low-energy range of \textit{XMM-Newton} and the higher-energy range of \textit{NuSTAR}, we can analyse all the reflection spectral features (because the Compton hump is visible only in the \textit{NuSTAR} spectrum because the \textit{Epic-pn} one is limited to 10 keV). Nevertheless, we left the continuum parameters free to vary among the two spectra and constrained only the inner disc radius, the ionisation parameter, and the inclination angle to be the same. This is because when the ionisation parameter is left free, the best fit gives compatible values for the two spectra. Moreover, we do not expect changes in the inner disc radius because the accretion rate in this source is quite stable over the years. We improved the fit by removing the partially covering component and by introducing a soft black-body (\textsc{bbody} in Xspec) with a temperature $kT_{bb}= 0.2$ keV.  We also added the reflection model, which is used to replace the broad Gaussian line at 6.7 keV, and kept the two Gaussian emission lines related to Fe I $K_{\alpha}$ and $K_{\beta}$, respectively. As done before, the inter-calibration constant was set to 1 for the Epic-pn and FPMA data sets, as the normalisation relative to the \textit{NuSTAR} spectra (as many other spectral parameters) are free and not fixed to those of \textit{XMM-Newton}. The RGS spectral parameters were linked to the pn parameters, except for the inter-calibration constant, which was left free to vary, and similarly for the FPMB spectrum with respect to the FPMA spectrum.

The first reflection model that we used includes the convolution model \textsc{Rfxconv} \citep[developed by][]{Kolem_2011}, which we used to reproduce only the reflection component. We applied to the continuum Comptonisation model \textsc{nthComp}, whose parameters are linked to those of the continuum \textsc{nthComp} that is already included in the model. The reflection component is smeared by the convolution model \textsc{Relconv} \citep{Dauser} to take Doppler and relativistic effects into account.
The \textsc{Relconv} model assumes a broken power law for the disc emissivity law, with two different indexes below and above a given radius R$_{br}$. We forced the two indexes to assume the same value because our spectra are not sensitive enough to reveal a difference in the emissivity index.

The second model includes the \textsc{diskline} model, which only describes the smeared iron emission line profile. This was combined with the \textsc{pexriv} model, which is used to represent only the reflection component, which  is an exponentially cutoff power-law spectrum reflected from ionised material \citep{Zdziarski}.
 The parameters of \textsc{pexriv} that characterise the continuum spectrum were appropriately linked to those of the \textsc{nthcomp} component: we set the electron temperature to 2.7 times that of the \textsc{nthcomp} (as is appropriate for a saturated Comptonisation, see also \citealt{Egron_2013}), and we linked the photon index of the power law in \textsc{pexriv} with the index of the Comptonisation component. In order to smear the spectrum for relativistic effects, we multiplied the reflection component by the convolution model \textsc{rdblur} (developed by \cite{Fabian_1989}).
To fit the residuals evident in the spectra at low energies (see Fig. \ref{fig:Iaria}), we added several Gaussian components in addition to the two that are related to the iron K-shell emission, which are interpreted as the emission lines associated with the atomic transitions shown in Table \ref{lines}. 
We also added an absorption edge at 9.44 keV, which can be traced back to the Fe XXVI K-edge that is necessary to fit the \textit{NuSTAR} residuals. Because this feature is not required to fit the \textit{XMM-Newton} spectrum, it may be a systematic feature, and therefore we froze its absorption depth to zero for the RGS and Epic-pn spectra.

We also tried to model the spectrum with other reflection models: with the self-consistent model \textsc{RelxillCp} \citep{Dauser_2016}, and with a variant of Model 2, where the component \textsc{pexriv} is replaced by \textsc{pexrav}. Model \textsc{RelxillCp} should predict a soft excess in the data that might also account for the \textsc{bbody} component; moreover, the \textsc{pexriv} component may not adequately describe the ionised reflection physics, and for this reason, we tried to replace it with \textsc{pexrav}, which assumes neutral matter in the disc.
However, the \textsc{RelxillCp} component proved unable to model the spectrum of this source adequately (the $\chi^{2}$/dof is 2193.0/1558 and 1636.3/1376 for the \textit{XMM-Netwon} and \textit{NuSTAR} data, respectively). The inadequacy of \textsc{RelxillCp} in describing our data might be related to the temperature of the seed photons, which is fixed at 0.06 keV in the model. This is significantly different from the temperature of 0.6-0.8 keV that we find for this system.
%\textsc{RellxilCp} includes a Comptonization spectrum with a temperature of the soft seed-photons fixed at 0.06 keV, which is in contrast with the seed-photon temperature that we find for this source (0.6-0.8 keV).
We added a low-energy exponential roll-off to the model using the \textsc{expabs} component to solve this problem, but without the \textsc{expabs} or \textsc{bbody} component the \textit{XMM-Newton} spectrum shows significant residuals at low energies (Fig. \ref{fig:Relxillcp}). The associated $\chi^{2}$ over d.o.f. is 3830.3/2935 and 3472.6/2932 for the model without and with the \textsc{bbody} component, respectively, and the best-fit parameters are reported in Tab. \ref{relxill}. 
The model including \textsc{pexrav} achieved a good fit to the data, with a lower $\chi^{2}$ than what we obtain with \textsc{pexriv} ($\chi^{2}$/dof = 3291.1/2929), with consistent best-fit values of the parameters when the inner disc radius is fixed to 75 R$_{g}$. When this parameter was left free to vary, however, the fit is unstable and the best-fit values are physically inconsistent. 
We therefore conclude that these models do not provide a better fit to the data in our case.
%Using the \textsc{RelxillCp} component not assured a better fit of the model to the data failing to fit the soft excess in the data,  while the \textsc{pexrav} one was unable to reach stable values of different parameters, like R$_{in}$ and inclination angle, even if when these parameters are fixed assured a better fit with a lower chi-sware. Thus, the use of \textsc{pexriv} component is preferable in this case,  since it can achieve a stable fit constraining the inner radius of the disc and the ionization parameter.
The results obtained by these fits are shown and discussed in Appendix A.

The shape of the relativistic iron line profile is mainly affected by different parameters that characterise the reflection models:  the inner disc radius R$_{in}$, the system inclination with respect to our line of sight,  the ionisation parameter $\xi$, and the reflection fraction defined as the fraction of the radiation reflected by the disc in terms of the solid angle $\Omega/2\pi$ that the disc subtends, as seen from the corona \citep{Kolem_2011}.
In our first fit session, we realised that the model that includes the \textsc{rfxconv} component was unable to constrain the value of the disc inner radius and the system inclination. Considering that the value of the system inclination has been discussed several times in the literature, we therefore decided to keep it frozen at the value found by \cite{2003Jonker},  82 degrees, while we kept the disc inner radius fixed at the magnetospheric radius expected for this source, that is, 75 R$_{g}$. We estimated this value as a fraction $\phi$ of the Alfvén radius R$_{A}$, defined by the relation given by \cite{King},
\begin{equation}\label{Alfven}
    R_{m}=\phi R_{A}= \phi \times 2.9 \cdot 10^{8} L_{37}^{-2/7} M_{1}^{1/7} R_{6}^{-2/7}\mu_{30}^{4/7} \; \mathrm{cm}
,\end{equation}
where M$_{1}$ is the mass of the NS in units of 1.4 M$_{\odot}$, R$_{6}$ is its radius in units of 10$^{6}$ cm, L$_{37}$ is the luminosity in units of 10$^{37}$ erg/s, and $\mu_{30}$ is the magnetic dipole moment in units of 10$^{30}$ G cm$^{3}$. The parameter $\phi$, set to $\sim$ 0.4, takes the case of accretion of matter through a geometrically thin disc into account, which reduces the size of the magnetosphere in the equatorial plane. Considering that \cite{Jonker_2001} have estimated a magnetic field of 8 $\times$ 10$^{10}$ G assuming a luminosity at the Eddington limit of 1.8 $\times$ 10$^{38}$ erg/sec, we obtained an inner radius of 149 km, that is, about 75 R$_{g}$ (see also \citealt{2015A&A...577A..63I}).

\begin{figure}
    \centering
    \includegraphics[width=.5\textwidth]{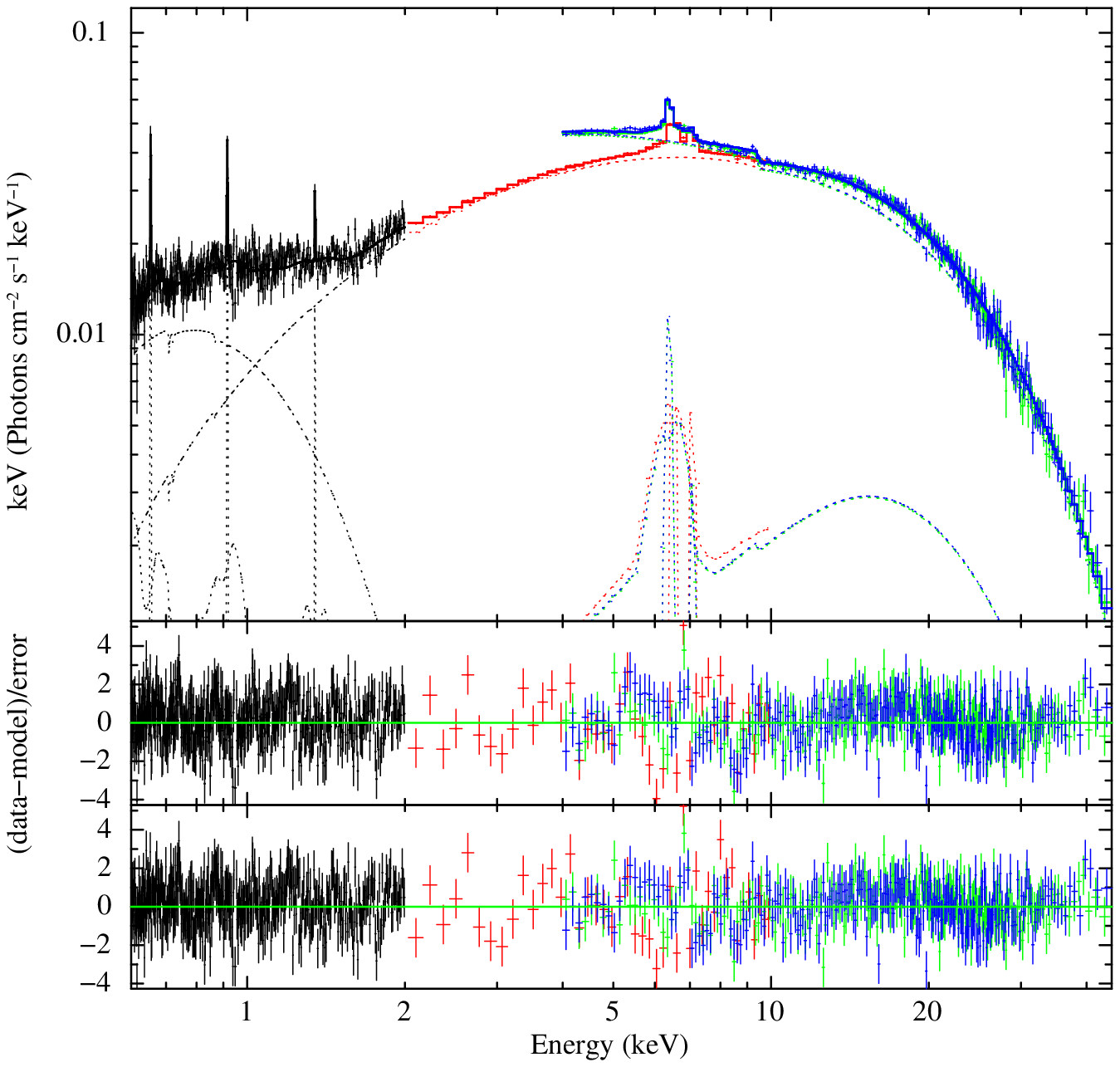}
    \includegraphics[width=.5\textwidth]{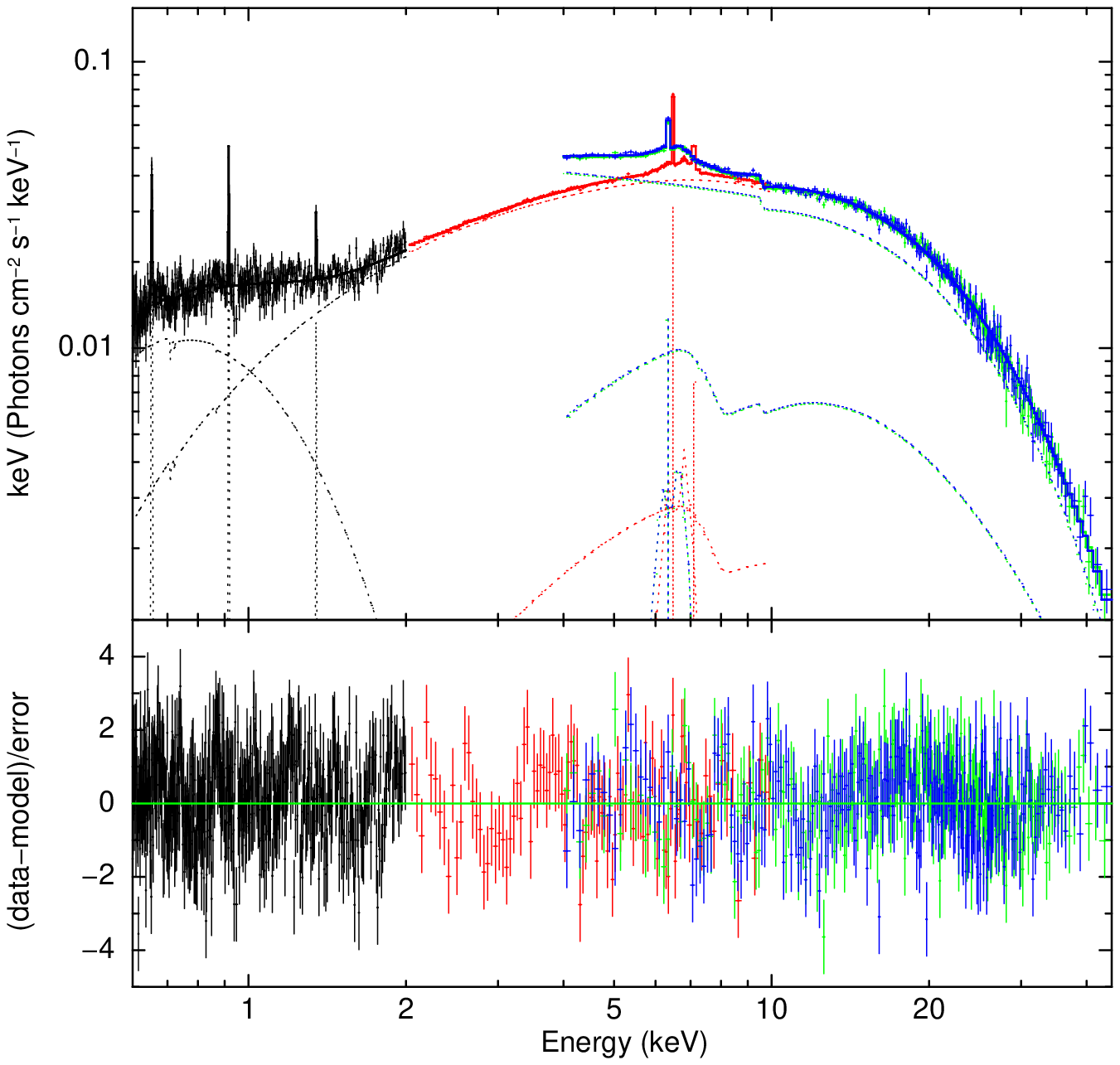}
    \caption{\label{Reflection_plot} Spectra and residuals in units of sigma with respect to the models described in the text. 
   The top panel shows the fit to the model $constant\cdot edge\cdot TBabs\cdot(6gaussian+ bbody + relconv\cdot rfxconv\cdot nthComp + nthComp)$ together with the residuals in units of sigma obtained by fixing the parameter R$_{in}$ to 75 R$_{g}$, and those obtained when R$_{in}$ assumes its best-fit value of 15 $R_g$. The bottom panel shows the fit to \textsc{Model 2}: $constant\cdot edge\cdot TBabs\cdot(6gaussian+ bbody + diskline + rdblur\cdot pexriv + nthComp)$ together with the residuals in units of sigma with respect to the model.}
\end{figure}

The best-fit values related to the models are reported in Table \ref{Reflection}:
the obtained parameters are consistent with the expected parameters, and the residuals (shown in Fig. \ref{Reflection_plot}) suggest a good fit of the data.  We tested the improvement of the fit using the statistical test \textsc{F-test} by comparing our model, which includes the self-consistent reflection component, with a model containing the same components except for the reflection, which was instead fitted with another broad Gaussian at 6.7 keV that was used to model the Fe XXV fluorescence line. We obtain a probability of chance improvement of $\sim 9 \times 10^{-34}$, meaning that the inclusion of the \textsc{RfxConv} significantly improves the quality of the fit with respect to a simple Gaussian component. 

We also fitted the data to a model that uses a combination of \textsc{pexriv} and \textsc{diskline} to fit the reflection component instead of \textsc{RfxConv}. Although this model is not self-consistent in the sense that the iron line is not self-consistently calculated from the reflection spectrum, we find that this provides a better fit with respect to \textsc{RfxConv},
%model that combines the \textsc{pexriv} and \textsc{diskline} components is not self consistent, 
as shown by the results reported in Table \ref{Reflection}. It clearly improves the $\chi^{2}$ value and allows obtaining a constraint both for the system inclination and for the inner disc radius, perfectly in line with what we expect for the system. We tested the improvement of the fit using the statistical test \textsc{F-test} that compared the two relativistic models, and we obtained a probability of chance improvement of $\sim 5.4 \times 10^{-22}$, meaning that \textsc{Model 2} ensures a significantly better fit to the data.

Keeping the inclination parameter fixed, we tried to let the inner radius free to vary between 6 and 200 $R_g$ in the \textsc{RfxConv} model, and we note  that the $\chi^{2}$ value improves significantly, ranging from 3430 with 2921 dof (degrees of freedom) to 3378 with 2920 (dof), reaching a best-fit value of $15 \pm 3\, R_g$ (see Table \ref{Reflection}). 
\begin{figure}
    \centering
    \includegraphics[width=.5\textwidth]{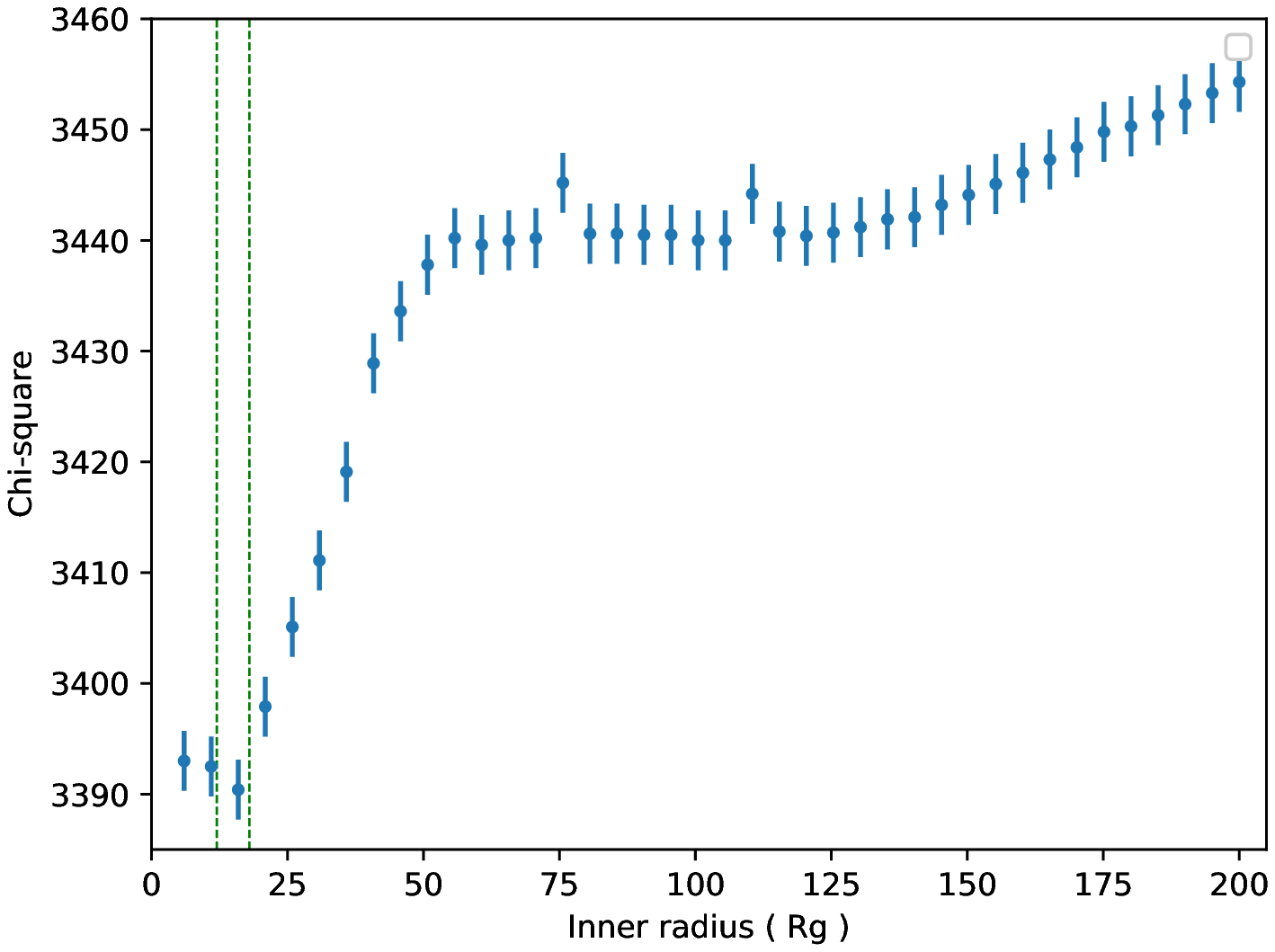}
        \includegraphics[width=.5\textwidth]{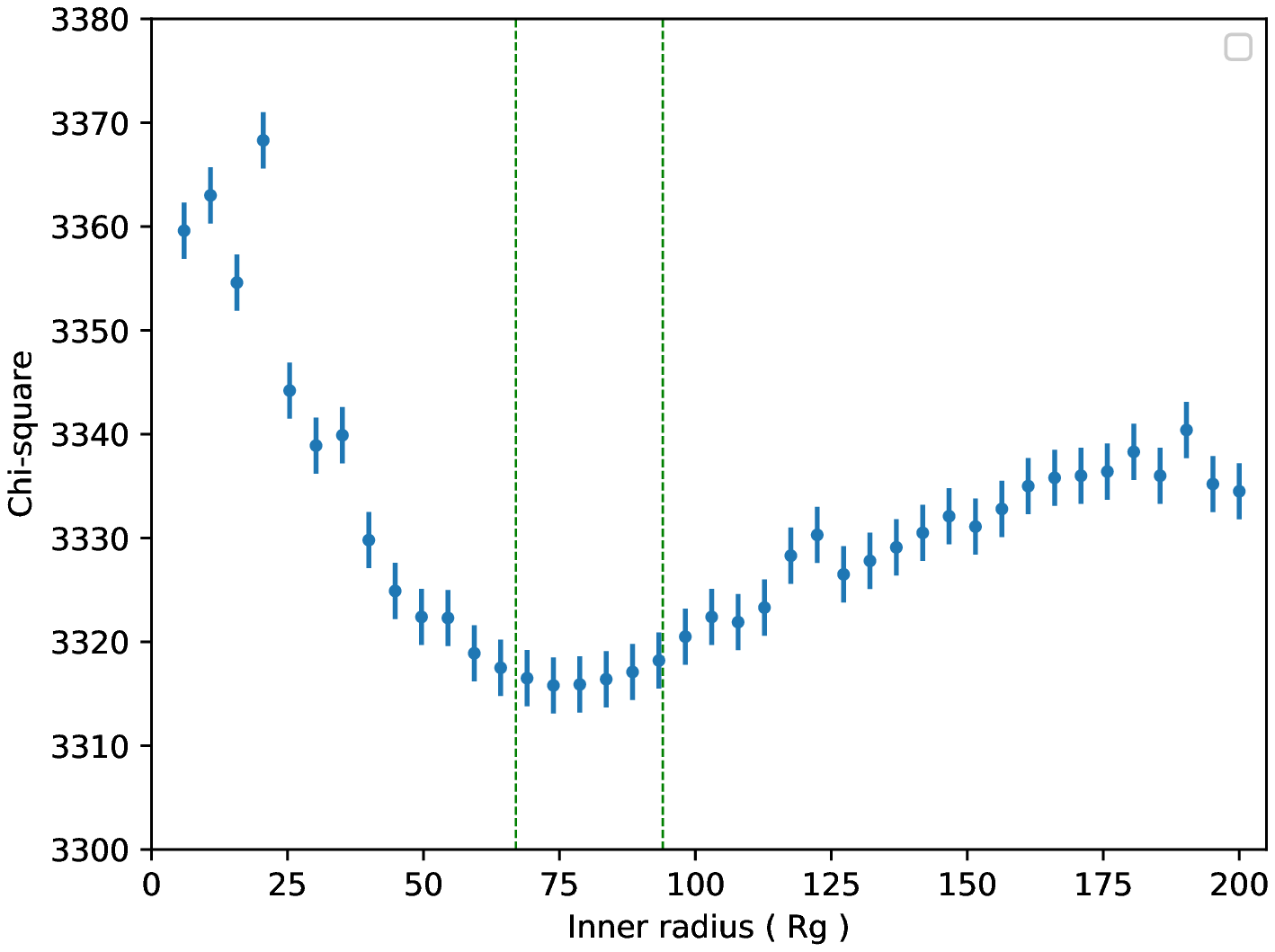}
    \caption{Variation in $\chi^{2}$ as a function of different values of the inner radius of the disc from 6 to 200 R$_{g}$ using \textsc{Model} 3 (top panel) and \textsc{Model} 2 (bottom panel). The highlighted boxes in the figures show the best-fit value of the inner disc radius for the two models (R$_{in}=15 \pm 3$ R$_{g}$ and R$_{in}=75^{+19}_{-7}$ R$_{g}$), and the error bars associated with each $\chi^{2}$ value represent $\Delta \chi^{2}=2.7,$ corresponding to the 90\% confidence level for a single parameter.}
    \label{fig:Steppar}
\end{figure}
To test the $\chi^{2}$ improvement, we used the Xspec command \textit{steppar} to perform a fit while stepping the value of the parameter through the range between 6 and 200 R$_{g}$, with 39 steps, using \textsc{Model} 3 and \textsc{Model} 2. The results related to the \textsc{rfxconv} model show that the $\chi^{2}$ reaches a minimum at the best-fit value and increases rapidly as the radius increases, achieving higher but flat values between 50 and 125 R$_{g}$, and then increasing again for higher values of the radius (Fig. \ref{fig:Steppar}). This result suggests a geometry of the system in which the inner region of the accretion disc is close to the NS surface, although the presence of coherent pulsating emission and the estimated value of the magnetospheric radius make this low value for the inner radius of the accretion disc unlikely. However, this high improvement in the fit quality cannot be neglected; for this reason, we discuss this result in the following section. It is also noteworthy that this result is in contrast with what we find using the \textsc{pexriv+diskline} to model the reflection spectrum because in this case, an absolute minimum of the $\chi^{2}$ is found for an inner disc radius in the range of R$_{in}=75^{+19}_{-7}$ R$_{g}$ (bottom panel of Fig. \ref{fig:Steppar}). 

\section{Discussion}

We analysed non-simultaneous \textit{XMM-Newton} and \textit{NuSTAR} observations of 4U 1822-371 by studying the average broad-band spectra and focusing on the possible presence of reflection features. As seen in the previous section, the addition of the reflection models improved the fit by reducing the residuals related to the iron fluorescence line and Compton reflection hump. The quality of our analysis was then ascertained using an F-test statistics, which returned a very low probability of chance improvement and confirmed the need of a reflection component, despite the high inclination of the system (note that the emission of the accretion disc decreases as the cosine of the inclination angle).
%Added by TeX Support    
\begin{table}[]\centering
 \caption{\label{lines}Emission lines at soft X-ray energies. }
 \small\addtolength{\tabcolsep}{-5pt}

 \begin{threeparttable}
\begin{tabular}{lcccc}
\toprule
\toprule
Id. & Expected&   E  & $\sigma$ &   N (10$^{-4}$)  \\
&(keV)&   (keV) &  (eV) &   (ph cm$^{-2}$ s$^{-1}$) \\
\toprule
\toprule
O VIII & 0.6536 & 0.6537$_{-0.0004}^{+0.0001}$      &  0.9$\pm$ 0.3   & 2.9 $\pm$ 0.3   \\
Ne IX (i) & 0.9162 &0.9149  $\pm$0.0007    & 1.6 $\pm$ 0.9 & 1.9 $\pm$ 0.3 \\
 
Mg  XI (i)& 1.3433 & 1.347$_{-0.003}^{+0.002}$         &  <5    & 0.8 $\pm$ 0.2  \\
 
Si XIV & 2.005 & 2.06  $\pm$ 0.06      & 18$^{+5}_{-3}$    & 2.0 $\pm$0.4   \\
 \hline
\end{tabular}
\end{threeparttable}
\end{table}
In the following, we first summarize the best-fit results, discussing the case in which the accretion disc extends close to the NS, and then we propose a possible scenario that can explain the observed features.    

%Added by TeX Support
\begin{table}[]\centering
 \caption{\label{Reflection} Best-fit values for the parameters of the two models described in the text.  }
\begin{threeparttable}

  \resizebox{0.5\textwidth}{!}{\begin{minipage}{\textwidth}

\begin{tabular}{llccc}
\toprule
Component & Parameter & Model 1\tnote{$\dagger$} & Model 2\tnote{$\star$} & Model 3\tnote{$\ddag$}  \\
\toprule 
%{\sc constant} & factor (RGS+FPMA) & 1 (frozen)& \\
%& factor (PN) & 1.04$_{-0.01}^{+0.02}$ \\
%& factor (FPMB) &  1.008$ \pm 0.003$ \\ 
\\
{\sc edge} & E(keV) & $9.44 \pm 0.08$ & $9.58_{-0.07}^{+0.08}$   &  9.50 $\pm 0.08$\\
 & $\tau$ & 0.075$\pm 0.008$ & $0.091 \pm 0.06$   & 0.065$^{0.006}_{-0.007}$ \\
 
{\sc tbabs  } & nH (10$^{22}$) & $0.15^{+0.02}_{-0.01}$ &  0.12\tnote{*}  & 0.168$^{+0.004}_{-0.005}$ \\

{\sc bbody} & kT(keV) &  $0.194^{+0.005}_{-0.006}$ &  $0.196 \pm 0.003$ & $0.194 \pm 0.002$  \\
 & N (10$^{-4}$) & $3.3^{+0.3}_{-0.2}$ &  $3.14 \pm 0.04$ & $3.46^{+0.03}_{-0.19}$ \\ \\
 
{\sc xmm } & $\Gamma$ & $1.334^{+0.016}_{-0.004}$ & $1.300 \pm 0.001$ & $1.394^{+0.002}_{-0.001}$  \\
{\sc nthComp} & kT$_{e}$(keV) & $3.31_{-0.06}^{+0.07}$& $3.26\pm 0.03$ & $3.79 \pm 0.05 $  \\
 & kT$_{bb}$(keV) & $0.84 \pm 0.01$ & $0.757 \pm 0.002$ & $0.914 \pm 0.002 $ \\
 &  N (10$^{-3}$) & $9.6 \pm 0.2$ & $10.2^{+0.1}_{-0.2}$ & $9.64^{+0.02}_{-0.15}$  \\ \\
 
{\sc nustar }& $\Gamma$ & 1.538 $\pm 0.005$ & $1.4841_{-0.0009}^{+0.0007}$ & 1.55 $\pm 0.001$   \\
{\sc nthComp}& kT$_{e}$(keV) & $5.02 \pm 0.03$ &  4.87 $\pm 0.02$  & $ 5.00 \pm 0.02 $    \\
 & kT$_{bb}$(keV) & $ 0.86_{-0.03}^{+0.02}$ & $0.557_{-0.002}^{+0.001}$ & $ 0.846_{-0.001}^{+0.002}$     \\
 &  N (10$^{-2}$) & $1.64 \pm 0.06$ &  2.57$_{-0.07}^{+0.08}$ &$ 1.671_{-0.025}^{+0.005}$    \\ \\ 
 
{\sc relconv} & Index & $2.6^{+0.3}_{-0.2}$& - & $2.16 \pm 0.06 $\\
 & Rbr & 250\tnote{*} & - &  250\tnote{*} \\
% & a & 0 (frozen) \\
 & Incl (deg) & 82\tnote{*} & - & 82\tnote{*} \\
 & R$_{in}$ & 75\tnote{$\dagger$}& - & $15 \pm 3$    \\
& R$_{out}$(10$^{3}$) & 1.0\tnote{*} & - & 1.0\tnote{*}  \\

 {\sc rfxconv} & rel$_{refl}$ & $-0.61^{+0.04}_{-0.05}$& - &  >-0.9 \\
% & Fe$_{abund}$ & 1 (frozen) \\
% & cosIncl & 0.139\\
 & log$_{xi}$ & $2.00^{+0.07}_{-0.21}$& - &  $1.72^{+0.03}_{-0.09}$ \\ \\

  {\sc diskline} &  E$_{XMM}$ (keV) & - & $6.58 \pm 0.03$ & - \\
&  E$_{NuSTAR}$ (keV) & - & $6.41 \pm 0.05$ & - \\
 & Index  & -& $-2.3 \pm 0.3$ & - \\
 & R$_{in}$ & - & $75_{-9}^{+11}$ & -\\
 & Incl (deg) & - &$85.7^{+0.1}_{0.4}$  & - \\
 & N$_{XMM}$ (10$^{-4}$) & -& $5.1^{+0.4}_{-0.2} $ & -\\ 
  & N$_{NuSTAR}$ (10$^{-4}$) & -& $5.0^{+0.7}_{-0.6}$ & -\\ 
{\sc pexriv}% & PhoIndex & 
% & foldE(keV) & 
 & rel$_{refl}$ & -&-1\tnote{*}&  - \\
% & Redshift & 0 (frozen) \\
% & abund & 1 (frozen) \\
 %& Fe$_{abund}$ & 1 (frozen) \\
  & xi & -& $9 \pm 2$ & - \\
 &  N$_{XMM}$& -& $0.125^{+0.008}_{-0.004}$ & -\\
 & N$_{NuSTAR}$ & -& $<0.51$  & -\\

\hline
\\
{\sc xmm } & E (keV) & $6.51  \pm 0.01$& $6.505^{+0.005}_{-0.02}$ & $6.51 \pm 0.01$  \\
{\sc gaussian} & $\sigma$ (eV) &<77* &  1.8E-14* & 44* \\
 &  N  (10$^{-4}$) & $3.4 \pm 0.3$&  $3.16 \pm 0.2$ & $3.4\pm 0.2$  \\
 \\
 & E (keV) & $7.08^{+0.02}_{-0.01}$ & $7.10 \pm 0.01$&  $7.06 ^{+0.02}_{-0.01}$\\
 &  $\sigma$ (eV) & $49^{+1}_{-3}$& 1.8E-14\tnote{*} & 44$^{1}_{-2}$ \\
 & norm (10$^{-4}$) & $2.4 \pm 0.2$ &  $1.4 \pm 0.2$ & $2.0 \pm 0.2$\\
 \\
 {\sc nustar} & E (keV) & $6.38 \pm 0.02$ &  $6.35 \pm 0.05$ & $6.38 \pm 0.02$ \\
{\sc gaussian} &  $\sigma$ (eV)  & 75\tnote{*}& 1.8E-14\tnote{*} & 75\tnote{*}  \\
 &  N  (10$^{-4}$) & $4.2 \pm 0.4$&  $2.4 \pm 0.4$ & 4.1$\pm 0.4$ \\ \\
 
 & E (keV)  & 7.08\tnote{*}&  7.11\tnote{*} & 7.06\tnote{*} \\
 &  $\sigma$ (eV) &  $75$\tnote{*} & 1.8E-14\tnote{*} & 75\tnote{*}\\
 &  N  (10$^{-4}$) & $0.9\pm 0.3$&$<0.1$ & $0.5 \pm 0.3$ \\ \\
\hline
 & $\chi^2/dof$ & 3430.23/2921  & 3303.95/2916 & 3377.9/2920     \\
 & $\chi^2_{red}$ & 1.17434 & 1.13304 & 1.15681    \\ \\ 
 \hline
\hline
\end{tabular}
\end{minipage}}
\begin{tablenotes}
\footnotesize
\item [*]Kept frozen during the fit.

\item[$\dagger$]\textsc{Model 1} : \textsc{constant*edge*TBabs*(6gaussian+ bbody + relconv*\\rfxconv*nthComp + nthComp)}.
\item[$\star$]\textsc{Model 2} : \textsc{constant* edge* TBabs*(6gaussian+ bbody + diskline + \\ rdblur*pexriv + nthComp)}.
\item[$\ddag$ ]\textsc{Model 3} has the same components of Model 1, but in this case the \\ inner radius R$_{in}$ is free to vary.
\end{tablenotes}
\end{threeparttable}
\end{table} 

The continuum emission is well fitted by the \textsc{bbody} and \textsc{nthComp} components. The first marks the presence of a black-body emission at 0.196 keV that may be produced by the disc or the NS surface. The saturated Comptonisation spectrum is produced in an optically thick corona with $\tau \sim 19 - 20$ with an electron temperature for Model 1 and Model 2 of 3.31$^{+0.07}_{-0.06}$ keV and 3.26 $\pm$ 0.03 keV, respectively, for the EPIC/pn plus RGS spectrum and 5.02$\pm 0.03$ keV and $4.87 \pm 0.02$ keV for the \textit{NuSTAR} spectra. The mismatch between the temperatures of the Comptonisation component may depend on the different epoch in which the two observations were performed. Although the two spectra show small but significant differences, we do not expect a significant variation in the reflection continuum spectrum because the latter contributes to a small fraction of the total spectrum.

%%%%%%%%%%%%%%%
To understand the geometry of the system, we first determine which region produces the continuum emission. We calculated the black-body radius for both models, that is, the radius of the (spherical) region from which the black-body radiation is emitted. Then we inferred the size of the (spherical) region producing the 
seed photons experiencing inverse Compton scattering off the hot electrons of the corona.
%obtained the seed photons radius, that is the size of the (spherical) emission region of the seed photons that experience inverse Compton scattering off the hot electrons of the Corona.
Matching the expression that defines the normalisation parameter in the \textsc{bbody} component with the Stefan-Boltzmann law, which links the radiation flux to the fourth power of the temperature, we have that
\begin{equation}
K=\frac{4\pi R_{bb}^{2}\sigma T^{4}}{10^{39} \text{erg}}\bigg(\frac{10 \text{kpc}}{D}\bigg)^{2}\; ,
\end{equation}
where $\sigma$ is the Stefan-Boltzmann constant that takes the value $\sigma$ = 5.6704 $\times$ 10$^{-5}$ erg/(cm$^{2}$s K$^{4}$), K is the black-body normalisation, which depends on the black-body observed flux, %. The temperature T is obtained by the fit but it should be expressed in Kelvin (1 keV= 1.16 \times 10$^{7}$ K) while
and D is the distance to the source, which was assumed to be 2.5 kpc (\citealt{Mason}). We did not take the error on the distance during the calculations of the parameters into account because \citet{Mason} reported the average value without an associated error. Taking these parameters into account, we find a black-body radius of 10.7 $\pm$ 0.3 km and 10.1 $\pm$       0.4 km for Model 1 and Model 2, respectively. Both radii have to be corrected by a hardening factor of $\sim 1.4$ because of distortions of the black-body spectrum mostly due to Thomson scattering (see e.g. \citealt{Foster1986}). This results in a black-body radius of 15.0 $\pm$ 0.4 km and 14.2 $\pm$ 0.6 km, respectively.

To determine the seed-photon emission radius, we analysed the best-fit values obtained for the Comptonisation spectrum and \textit{\textup{de-Comptonised}} it to trace the continuum emission before it interacted with the corona. %We first have created an artificial response matrix in Xspec, from which we took 
From the two models, we extrapolated the value of the flux, F$_{x}$ = (1.65 - 0.92) $\times$ 10$^{-9}$ erg/s cm$^{-2}$, and calculated the bolometric luminosity. We can derive the optical depth of the corona using the relation of \cite{Zdziarski},
\begin{equation}
\Gamma +\frac{1}{2}=\left[\frac{9}{4}+\frac{1}{\left(k T / m_{e} c^{2}\right) \tau(1+\tau / 3)}\right]^{1 / 2}.
\end{equation}
Using the best-fit values of the parameters, we obtained values of the optical depth of $\tau$ =19 $\pm$ 1     and 20.4 $\pm$ 0.2 for the two models. Hence we can state that the Comptonisation spectrum is emitted by an optically thick corona surrounding the region of the system in which the seed photons are emitted, as we expect from the fact that we found a saturated Comptonisation spectrum, with a shape quite similar to that of a black-body, with low values of the photon index and electron temperature. %%, and a high optical depth.

In order to obtain the flux associated with the de-Comptonised spectrum \textit{f$_{seed}$}, we divided the bolometric flux of the Comptonisation component by the factor (1+\textit{y}), where \textit{y} is the Compton y-parameter calculated under the assumption of the optically thick case (\textit{y} = 9(1) - 10.6(2)), and we finally derived the seed-photon radius R$_{sp}$ from the following equation:
\begin{equation}
    R_{sp}=3\times 10^{4}\;f_{seed}^{1/2}\;T_{bb}^{-2}(\text{keV})\;D \;\; \text{km} \; ,
\end{equation}
obtaining a value of $0.9 \pm 0.1$ km and $1.16 \pm 0.02 $ km for Models 1 and 2, respectively. 

%Given the values obtained for the seed-photon radius, we propose the following associations: 
Because we found a black-body radius of 15.0 $\pm$ 0.4 km and 14.2 $\pm$ 0.6 km, which is compatible with the NS radius, it would be natural to associate the black-body emission region with the NS surface, while the value found for the seed-photon radius suggests an emission from a smaller region, such as the magnetic polar caps of the NS. The latter hypothesis may be justified by the coherent pulsations at 0.59 s detected by \cite{Jonker_2001}.
However, it is difficult to think that these two thermal emissions might originate in the compact object because the temperatures of the two components are very different (the seed-photon temperature is about 0.9 keV, and the black-body temperature is 0.2 keV) and because we would expect some (thermal black-body-like) contribution from the accretion disc (see below).

The value we found for the emissivity-law index for the reflection component agreeswith what we expect: we obtained a value of $2.6^{+0.3}_{-0.1}$ and $2.3 \pm$ 0.3 for Model 1 and Model 2, respectively. We expect an emissivity index of about 2-3 depending on whether the reflection spectrum is dominated by the illuminating flux coming from the central source
%%emission comes from the central object 
(in this case, the flux illuminating the disc decreases with distance as r$^{-2}$) or by the intrinsic disc emission (whose intensity decreases with distance approximately as r$^{-3}$). We found a value of the ionisation parameter $\xi$ (defined as $\xi=4\pi F_{x}/n_{e}$ \cite{Fabian_2000}, where F$_{x}$ is the X-ray flux and n$_{e}$ is the electron number density,) of 9 $\pm$ 2 and a $\log{\xi}$ of 2.00$^{+0.07}_{-0.21}$ for the two models, respectively, which is in line with the energies of the emission lines observed in the spectrum. This ionisation parameter implies a relatively low ionisation state of matter in the disc; the iron line is dominated by the Fe I-XX K$\alpha$ transition \citep{Fabian_2000}. 

As mentioned in the previous section, we found two different minima of the $\chi^{2}$ for Model 1, which differ for the values of the inner accretion disc radius. In the first case, the radius is truncated at the magnetospheric radius predicted for this pulsar, that is, 75 R$_g$, while for the second case, we obtained a configuration in which the disc extends closer to the NS (with an inner radius of 15 R$_g$), corresponding to a significantly lower $\chi^{2}$ value. 
When we used Model 2 to fit the spectrum, we find a clear minimum of the $\chi^{2}$ for a value for the inner disc radius of about $\sim 75$ R$_g$, compatible with the estimated position of the magnetospheric radius of the source. 
In correspondence, we find different values of the reflection fraction: when the radius is kept frozen at 75 R$_g$, this parameter assumes the stable value of 0.61$^{+0.04}_{-0.05}$, while when we let  the inner radius free, the reflection fraction tends to the value of 0.9, suggesting that when the accretion disc approaches the central object, the portion of the radiation reflected by the disc has to increase. The reflection fraction of Model 2 is instead fixed at -1: we cannot simultaneously determine the reflection fraction and the normalisation of the \textsc{pexriv} component because its normalisation cannot be fixed to that of the Comptonisation spectrum. These parameters are defined in different ways and cannot be easily compared. Thus we decided to let the \textsc{pexriv} normalisation free and fixed the reflection fraction at its maximum expected value. 
However, in order to give an estimation of the reflection fraction in this case, we evaluated the flux of the reflection component (\textsc{rdblur*pexriv+diskline}) and the flux of the Comptonisation continuum both in the energy range $0.1-100$ keV in order to calculate their ratio. The reflection component corresponds to $\sim 7\%$ of the Comptonisation component. 

In the following, we discuss the results described above considering a scenario in which the intrinsic luminosity of the central source is at the Eddington limit and most of this emission is blocked by the outer rim of the accretion disc. In this case, what we see is just the fraction of the intrinsic emission of the central source that is scattered along our line of sight by an extended optically thin corona.

%%propose a possible scenario that can explain the above discussion about the emission region of the continuum spectrum, the presence of the weak relativistic reflection component and the mismatch in the observed luminosity discussed in literature.

\subsection{Two coronae}

%In order to develop our hypothesis, we start discussing the source luminosity. 
The results discussed above are obtained considering the observed source luminosity of $\sim 10^{36}$ erg s$^{-1}$. However, this luminosity may not be the intrinsic source luminosity. Three independent arguments are present in  the literature, suggesting that the observed luminosity does not reflect the intrinsic luminosity emitted by the source:
i)  \cite{Burderi_2010} studied the orbital period evolution of the system and found a constraint on the orbital period derivative of $\dot{P}$ = 1.50(7) $\times$ 10$^{-10}$ s s$^{-1}$, which is three orders of magnitude larger than that calculated for a system assuming a conservative mass transfer. This constraint led the authors (see also \citealt{Bayless2010, Iaria2011, MAzzola})
to suggest a scenario in which the mass transfer rate from the companion is more than three times the Eddington limit (i.e. 1.8 $\times$ 10$^{-8}$ M$_{\odot}$/yr for a NS of 1.4 M$_{\odot}$), thus proposing a non-conservative mass-transfer scenario, where the mass in excess is expelled from the system.
ii) \cite{Hellier_Mason_1989} fitted an EXOSAT spectrum of 4U 1822-371 and found a ratio of the X-ray luminosity to the optical one of L$_{x}$/L$_{o} \sim 20$, a factor 50 lower than the typical value usually found for LMXBs, which might indicate that the observed luminosity is reduced by at least a factor 50 compared to the intrinsic one.
iii) \cite{Jonker_2001} found a pulse period derivative for 4U 1822-371 of (-2.85 $\pm$ \;0.04) $\times$ 10$^{-12}$ s s$^{-1}$, concluding that the source is spinning up \citep[see also][]{MAzzola,2015A&A...577A..63I,Jain_2010}.
Using the relation between observed luminosity and spin-up rate given by \cite{Ghosh_1979}, the authors find that for a luminosity of 10$^{36}$ erg/s, the NS magnetic field should be about $8 \times 10^{16}$G, which is extremely odd for an NS in a LMXB, while for a luminosity of 10$^{38}$ erg/s, around the Eddington limit, the expected value for the magnetic field is a more conceivable value of $8 \times 10^{10}$G.

Notwithstanding the high inclination angle of the system, we observe not only X-ray pulsations coming from the NS, but also a weak black-body spectrum that seems to be emitted from the NS surface, as well as signatures of a reflection component such as a Compton hump at high energy (10-30 keV) in the \textit{NuSTAR} spectrum and a broad Fe emission line with a profile compatible with a smearing produced by Doppler effects in the inner accretion disc.
A possible way to justify these pieces of evidence is to introduce an extended optically thin corona above the whole system that scatters only 1\% of the intrinsic luminosity emitted by the central source into the line of sight \citep[as proposed by][]{Iaria}. The consequence of this hypothesis is that all the flux that we observe is the part of the total emission that is produced by the source that is scattered along the line of sight.  If the extended Corona is optically thin, so that $\tau_{c}$ << 1, the scattered flux would be \citep{Iaria}
\begin{equation}
    F_{Obs}= L_{0}\frac{\left(1-\mathrm{e}^{-\tau_{c}}\right)}{4 \pi D^{2}} \simeq \frac{L_{0} \tau_{\mathrm{C}}}{4 \pi D^{2}}\; .
\end{equation}
Assuming a distance to the source of \textit{D}= 2.5 kpc, to obtain an intrinsic luminosity of 10$^{38}$ erg/s (Eddington limit) with an observed one of 10$^{36}$ erg/s, $\tau_{c}$ must be  $\sim$ 0.01. The presence of a tenuous corona around the system can also explain the values found for the black-body radius and seed-photon radius. In deriving the value of the two radii, we considered the observed luminosity of 10$^{36}$ erg/s, while we should have used the intrinsic luminosity that is probably two orders of magnitude higher. Because the luminosity is proportional to R$^{2}$, we must multiply the previous values by a factor 10  to determine the correct radii. With this, we obtain more reasonable values of R$_{bb}$ = 140 km and  R$_{sp}$ = 11 km, respectively. In this case, we can identify the black-body emission as due to the inner rim of the accretion disc, which is truncated by the magnetosphere at $\sim 140$ km from the NS centre, and the seed-photon emission region as the NS surface. This interpretation agrees with the fact that the black-body temperature is lower than the seed-photon temperature because the temperature in the system is expected to decrease at larger distances from the compact object. 
%%we expect the disc to be colder than the NS surface, in agreement with what we observe.
In this case, the best-fit value of the inner radius that is found using the reflection component \textsc{rfxconv} of $15 \pm 3$ R$_{g}$ is hard to reconcile with a black-body emission produced at 140 km (i.e. 70 Rg) from the NS.

%Considering that we obtained a best-fit value of the inner radius of the reflection component \textsc{rfxconv} of $15 \pm 3$ R$_{g}$, a blackbody emission produced at 140 km (i.e. 70 Rg) from the star means that this component may be ascribed to the inner accretion disc radius. Instead, an R$_{sp}$= 11 km implies that the seed photons of the Comptonized component may come from the NS surface. This interpretation also explains why the blackbody temperature is colder than the seed photons one: since the temperature in the system decreases at larger distances from the compact object we expect the disc to be colder than the NS surface, in agreement with what we observe.

\subsection{Inner accretion disc radius}

\begin{figure}
    \centering
    \includegraphics[width=.5\textwidth]{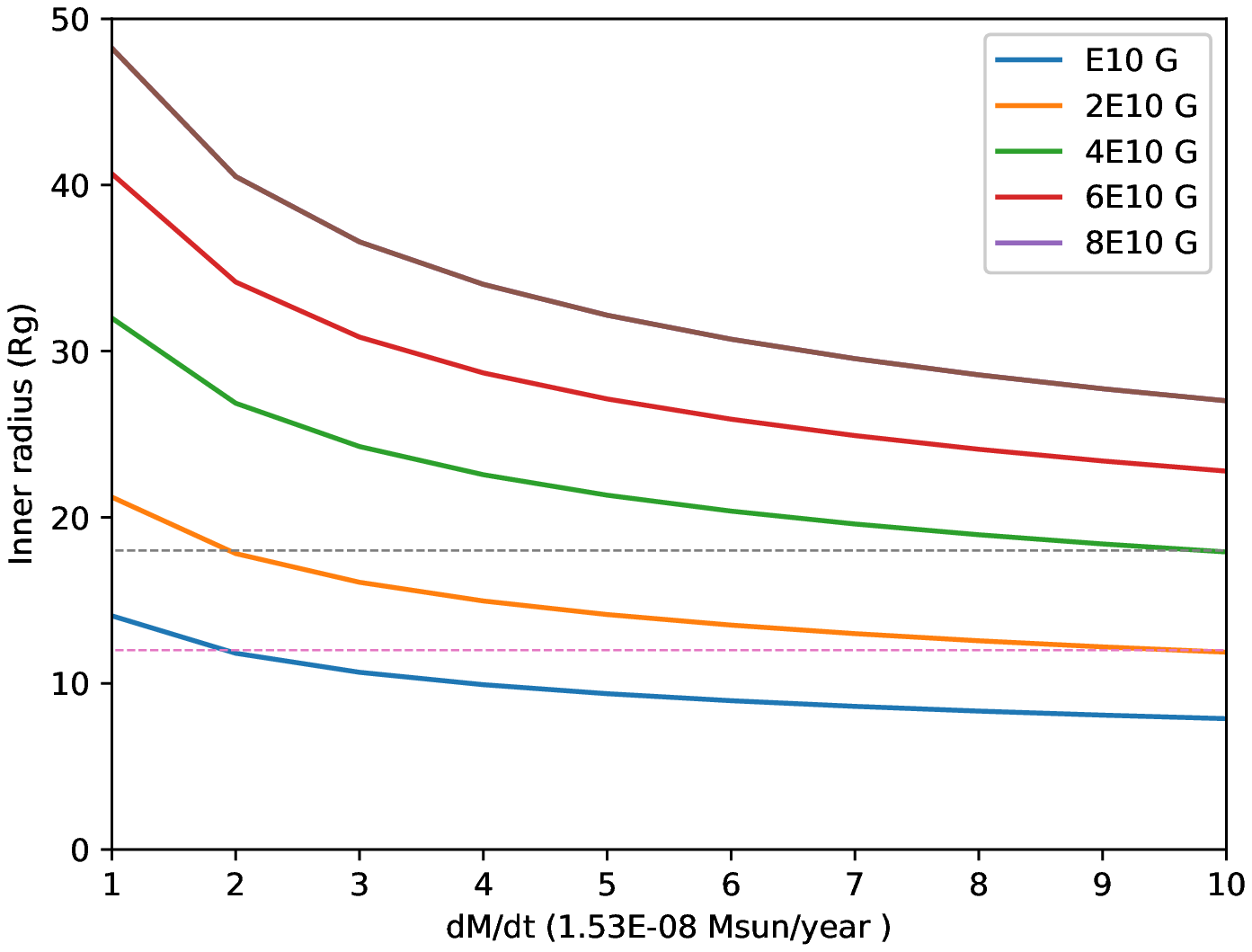}
\includegraphics[width=.5\textwidth]{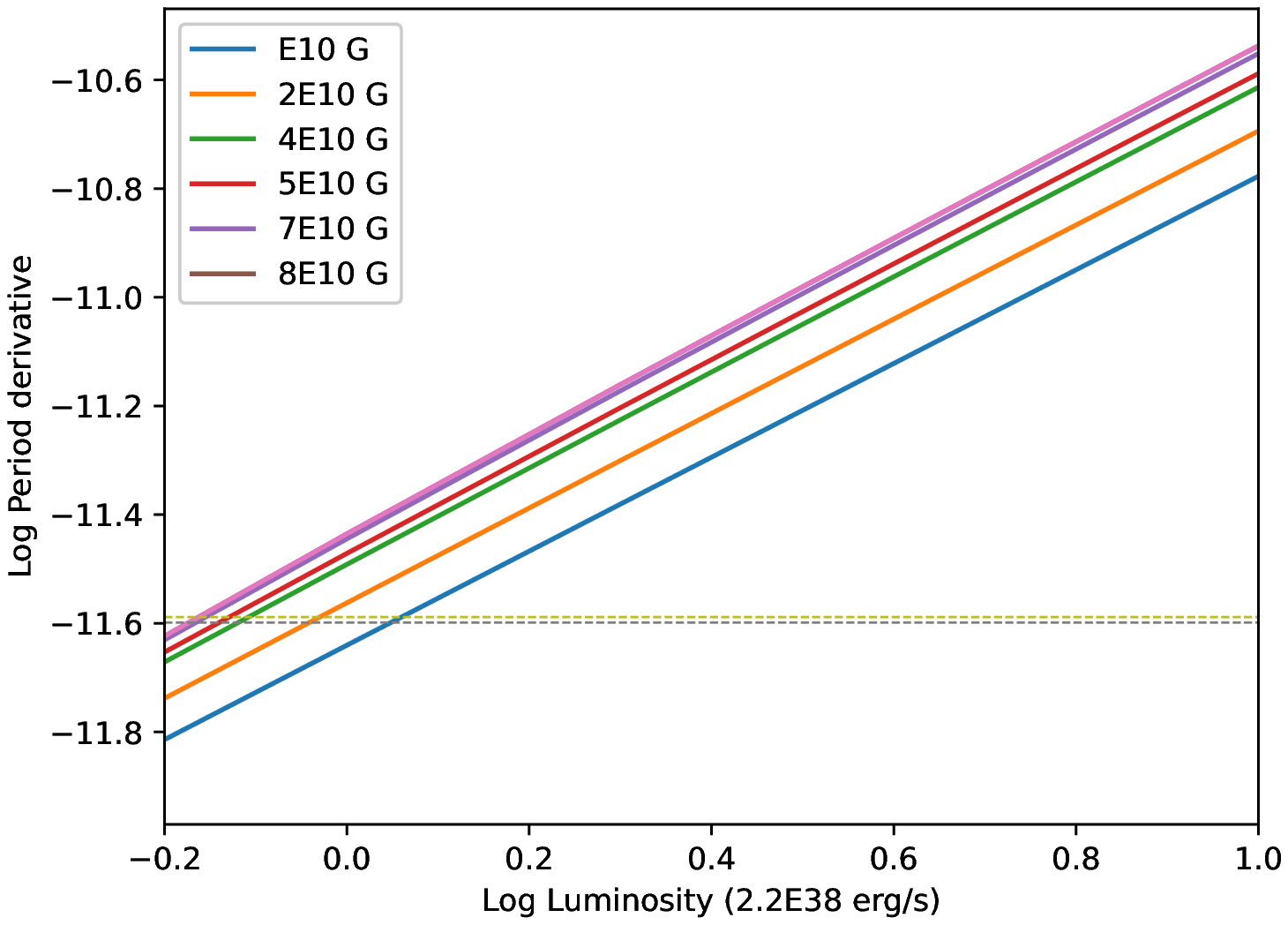}
    \caption{Inner radius of the accretion disc plotted vs. the intrinsic source luminosity for different values of the magnetic field (solid lines), obtained with  Eq. \ref{Alfv} (top panel). Derivative of the spin period as a function of luminosity for different values of the magnetic field on a logarithmic scale as indicated in the labels (bottom panel). The dotted lines shown in the plot are the logarithmic of $-2.595(11) \times 10^{-12}$ s/s, i.e. the spin period derivative reported in the literature \citep{MAzzola}.}
    \label{fig:R_ghosh}
\end{figure}

In the following, we show that a value for the inner disc radius as low as $\sim 15\, R_g$ may be obtained for this source  provided that the magnetic field is not stronger than $4 \times 10^{10}$ G.
To understand this, it is necessary to study the variation of the inner accretion disc radius of the system with respect to its mass accretion rate and magnetic field. For this purpose, we assumed  a NS mass of 1.69 M$_{\odot}$, as estimated by \cite{Iaria2011} for the 4U 1822-371 system, and a NS radius of R = 10$^{6}$ cm. We calculated the inner disc radius by varying the mass accretion rate in a range  1-10 times the Eddington limit (which is assumed to be 1.53 $\times$ 10$^{-8}$ M$_{\odot}$/yr for a NS with a mass of 1.69 M$_{\odot}$) and the magnetic field in a range of 1-8 $\times$ 10$^{10}$ G, using the expression reported by \cite{Sanna_2017},
\begin{equation}\label{Alfv}
    R_{\mathrm{m}}=\phi\;R_{\Lambda}=\phi\;(2 G M)^{-1 / 7} \mu^{4 / 7} \dot{M}^{-2 / 7}\; ,
\end{equation}
where G is the gravitational constant, M is the NS mass assumed to be 1.69 M$_{\odot}$, $\mu$ the star’s magnetic dipole moment, and $\dot{M}$ is the mass accretion rate. The factor $\phi$ depends on the structure of the accretion disc, but we can estimate it with the following expression \cite{Sanna_2017}:
\begin{equation}
    \phi=0.315 \;\kappa_{0.615}^{8 / 27}\; \alpha^{4 / 15}\; \mu_{26}^{4 / 189}\; \dot{M}_{-9}^{32 / 945}\; m^{76 / 189} \; ,
\end{equation}
where $\mu_{26}$, $R_{6}$, $\dot{M}_{-9}$ and \textit{m} are the NS magnetic moment in units of 10$^{26}$ G  cm$^{3}$, the NS  radius in units of 10$^{6}$ cm, the mass accretion rate in unit of 10$^{-9}$ M$_{\odot}$/yr, and the NS mass in units of solar mass. The mean molecular weight $\kappa$ can be assumed equal to 0.615 for fully ionised matter, while the parameter $\alpha$ that expresses the disc viscosity in the standard Shakura–Sunyaev model is usually set equal to 0.1 \citep{1973Shakura_Sunyaev}. The results are plotted in the top panel of Fig. \ref{fig:R_ghosh}: the region between the dotted lines at 12 and 18 R$_{g}$ represents the range of our best-fit value. It is easy to understand that this value can be found with a different combination of $\dot M$ and B, for example assuming an accretion rate of few times Eddington and a magnetic field with a strength of few 10$^{10}$ G.

In order to verify whether these values may be plausible for  4U 1822-371, we used the \cite{Ghosh_1979} equation, which links the spin-period derivative of the NS, its magnetic field and its luminosity, in order to understand which pairs of values return the spin period derivative reported in the literature for this source, 
\begin{equation}
-\dot{P}=5.0 \times 10^{-5} \; \mu_{30}^{2 / 7}\; n\left(\omega_{s}\right)\;M_{1}^{-3/7}\; I_{45}^{-1}\left(P L_{37}^{3 / 7}\right)^{2} \mathrm{s}/ \mathrm{yr}.\end{equation}
In this relation, M$_{1}$ is the NS mass in units of 1 M$_{\odot}$, $I_{45}$ is the NS moment of inertia in units of 10$^{45}$ g cm$^{2}$ , and P is the spin period of the source, that is, 0.5915669 s \citep{MAzzola}. The parameter n($\omega_{s}$) is the dimensionless accretion torque that is a function of the fastness parameter $\omega_{s}$. When $\omega_{s}$ < 0.95, we can use the following approximate expressions  \citep{Ghosh_1979}:
\begin{align}
\begin{split}
&n\approx 1.39\left\{1-\omega_{s}\left[4.03\left(1-\omega_{s}\right)^{0.173}-0.878\right]\right\}\left(1-\omega_{s}\right)^{-1} \; , \\ &
\omega_{s} \approx 1.35\; \mu_{30}^{6 / 7} M_{1}^{-2/7} \left(P L_{37}^{3 / 7}\right)^{-1} \; .
\end{split}
\end{align}
The values of luminosity and magnetic field used in the equations above vary in a range from 0.5 to 10 times the Eddington value and from 10$^{10}$ to 8 $\times$ 10$^{10}$ G, respectively. The results are shown in the bottom panel of Fig. \ref{fig:R_ghosh}, where the dotted line represents the derivative of the spin period measured for the source, (-2.59$\pm$ 0.03) $\times$ 10$^{-12}$ s/s \citep[e.g.][]{MAzzola}. Fig. \ref{fig:R_ghosh} shows that a luminosity and a magnetic field of the order of those found with the Eq. \ref{Alfv} are compatible with $\dot P$ relative to the source. In other words, we may obtain a magnetospheric radius as small as $\sim 15\, R_g$ assuming an intrinsic luminosity of a few times the Eddington limit and a magnetic field strength below $4 \times 10^{10}\, G$. 
%, meaning that the inner radius obtained by the best fit could be physically reliable for this source. On the other hand, also the  $\dot P$ found by assuming the expected values of luminosity and a magnetic field intensity (L= 10$^{38}$ erg/s and B = $8 \times 10^{10}$ G), not significantly deviate from the one reported in literature.

However, our discussion cannot fully justify this small inner accretion disc radius, which would be incompatible with the inner disc radius we infer from the black-body component (see above). Furthermore, although the assumed intrinsic luminosity may be plausible, the magnetic field value reported by several authors in the literature is $ \sim 8 \times 10^{10}$G. This value has been reported, for instance, by \cite{2015A&A...577A..63I}, who detected a possible cyclotron absorption feature at $\sim 0.7$ keV in the low-energy  spectrum of the source. On the other hand, the best-fit value constrained with \textsc{Model 2} (R$_{in}$ = $75_{-7}^{+19}$ R$_{g}$),  which ensures the best fit to our data with the lowest $\chi^{2}$, is perfectly in line with the value of luminosity and magnetic field expected for this system.

Alternatively, the low value indicated by the best-fit parameters of the reflection component \textsc{rfxconv}, if confirmed, might be explained by an excess of smearing of the iron line. It is possible that although the line is actually emitted at $45-75$ $R_g$, it appears more distorted than it should be, resulting in a lower inner disc value. The reason for this distortion is unclear (stronger Compton broadening than expected in the reflection models, blending of lines produced by different ions of iron, disc winds with high turbulent velocities), and may be connected to the accretion rate at the Eddington limit and how it changes the geometry of the disc that is viewed at this high inclination angle. 
Otherwise, considering that the best fit is obtained by separately fitting the iron line and the remaining reflection component, this result may be caused by a poor consistency between the line and the other reflection features, which may be due to the edge-on view and the fact that we do not directly observe the inner region, but just observe its emission scattered into the line of sight by the optically thin corona. The model with \textsc{rfxconv} returns a warning related to the slope of the power law, that exceeds the tabulated range, suggesting again that more suitable models are required to describe the reflection component of sources such as 4U 1822-371.

\subsection{Emission lines at low energies}
As reported in Tab. \ref{lines}, we detected four emission lines at soft X-rays
 that can be associated with emission from O VII, Ne IX, Mg XI, and Si XIV ions. We identified the Ne and Mg lines with the inter-combination lines, while the O and Si lines can easily be associated with the Lyman-alpha transition. To prove the improvement of the $\chi^{2}$ to the introduction of the four Gaussian lines, we used the F-test statistic. This method generally works when nested models are compared, and in principle, it should not be used to verify the addition of a Gaussian component in the model because the boundary of the possible values of the normalisation parameter is the null value (\cite{Protassov_2002}). However, when we extend the range of possible normalisation values to negative values, the F-test can be adopted to prove the improvement of the fit. We compared the best fit obtained with \textsc{Model 2} with a fit in which we individually removed each of the Gaussian lines. We find that the addition of all the Gaussian lines has a significance higher than 6$\sigma$. Because of the expected emission lines at these energies, we can therefore state that these features are most probably real and not related to systematic features. In addition to the F-test, another method to ascertain the significance of the lines is considering their normalisation and comparing the latter with the error, at 1 sigma, associated with the parameter. In this case, the significance of the lines is also higher than 6 sigmas for the emission lines associated with O VII, Ne IX, and Mg XI, and higher than 3 sigmas for the line associated with Si XIV.
Because the lines  do not show evidence of Doppler effects, we propose that they are emitted from an outer region of the disc (in accordance with \cite{Iaria2011}). However, further investigations are needed to accurately determine the emission region.

\section{Conclusions}
We have analysed \textit{XMM-Newton} and \textit{NuSTAR} observations of 4U 1822-371 with the aim of highlighting the presence of a reflection component in the spectrum, despite the high inclination angle of the system.
We first tried to reproduce the result obtained by \cite{Iaria}, who fitted the source spectrum without any reflection component, but with respect to this model, we obtained residuals at the iron line region and an excess of emission above $\sim 10$ keV, reminiscent of a Compton hump caused by a reflection component. We tried to fit the complex residuals at the iron line region with three Gaussian lines, but achieved unsatisfactory results. Therefore we tested two reflection models, \textsc{rfxconv} and \textsc{diskline + pexriv}, to fit the spectrum. We obtained a significant improvement in the fit, suggesting a detection of this component at the 7 sigma confidence level. To our knowledge, no reflection component in a source with this high inclination has been observed before because reflection should be very faint and difficult to detect in a source with an inclination angle higher than 80 degrees.
The main continuum emission is well fitted by a thermal black-body emission, with a temperature of 0.2 keV, and a satured Comptonised component with an electron temperature of $3.26 - 4.87$ keV, generated by the inverse Compton scattering of seed photons with a temperature of 0.8 keV in an optically thick corona ($\tau=19-20$), located in the innermost region of the system, possibly around the NS. 
We observe several emission lines that can be associated with O VIII, Ne IX, Mg XI, and Si XIV in the RGS spectrum. These lines may be produced in the bulge in the outer region of the accretion disc, as previously proposed \citep[e.g.][]{Iaria2011}, although a detailed orbital phase-resolved spectral analysis is required to corroborate this hypothesis \citep{Somero_2012}.

We updated the orbital ephemeris of the source by  adding two eclipse arrival times derived from \textit{NuSTAR} (2018) and Swift (2007) observations to those reported by \cite{MAzzola}. We fitted the delays of the eclipse arrival times with a quadratic, a cubic, and a sinusoidal model. We proved that by adopting the latter two instead of the quadratic model,  the fit significantly improves. In this way, we found an orbital period $P_{orb}$ = 5.57063023(34) hr and a $\dot{P}_{orb}$ value of 1.51(5) $\times$ 10$^{-10}$ s s$^{-1}$.  
In line with other results present in the literature, we assumed  that the source is emitting at an intrinsic Eddington luminosity of $2 \times 10^{38}$ erg/s, even though the observed luminosity is just 10$^{36}$ erg/s.
To explain the discrepancy between the intrinsic and the observed luminosity, as well as the presence of a reflection component in the spectrum, an extended optically thin corona ($\tau \sim$ 0.01) around the whole system has been suggested, which would scatter about 1\% of the intrinsic luminosity produced by the central source along the line of sight. Because the whole emission is scattered, this mechanism enables us to observe a reflection component, which otherwise would be very difficult to detect at such a high inclination angle. 

In our analysis, we found a singular feature related to the inner accretion disc radius: the best-fit value of the inner radius achieved with the relativistic component \textsc{rfxconv} is $15 \pm 3\, R_g$, with a clear improvement of the $\chi^{2}$ over the fit in which the inner radius was fixed to 75 R$_g$, which is the theoretically predicted value of the magnetospheric radius in this source. We discussed the possibility that the iron line is really emitted in a disc region so close to the compact object and concluded that this hypothesis at the Eddington luminosity requires a magnetic field of 10$^{10}$ G, which deviates from the value reported in the literature for this source (of 8.8 $\times$ 10$^{10}$ \citealt{2015A&A...577A..63I}). This result would also be at odds with the inferred radius of the black-body emission region of $\sim 140$ km that we identify with the inner rim of the accretion disc. Alternatively, the iron line might be emitted at the predicted magnetospheric radius, but the line profile appears to be more smeared than it should be, resulting in a lower value of the inner radius. The mechanism(s) originating this line distortion are unclear and might be related to the accretion rate at the Eddington limit and to the scattering the line photons undergo in their travel along the line of sight. However, using a combination of \textsc{diskline} and \textsc{pexriv} to fit the reflection component, we obtained an inner disc radius of $\sim 75\, R_g$ and a (slightly) better quality of the fit, showing that this value is still model dependent.  
Future high-statistics broad-band observations may 
%%allow an (orbital) phase-resolved spectral analysis and may
help to solve this ambiguity. 
\begin{acknowledgements}
The authors acknowledge financial contribution from the agreement ASI-INAF n.2017-14-H.0 from INAF mainstream (PI: A. De Rosa), and from the HERMES project financed by the Italian Space Agency (ASI) Agreement n. 2016/13 U.O and from the ASI-INAF Accordo Attuativo HERMES Technologic Pathfinder n. 2018-10-H.1-2020. We also acknowledge support from the European Union Horizon 2020 Research and Innovation Framework Programme under grant agreement HERMES-Scientific Pathfinder n. 821896. RI and TDS acknowledge the research grant iPeska (PI: Andrea Possenti) funded under the INAF national call Prin-SKA/CTA approved with the Presidential Decree 70/2016. RI acknowledges financial contribution from the agreement ASI-INAF n.2017-14-H.0, from INAF mainstream (PI: T. Belloni). 
\end{acknowledgements}
\bibliography{Bibliografia}
\bibliographystyle{aa}

 \begin{appendix}
\section{Another reflection component}

%Added by TeX Support
\begin{table*}[t]
 \caption{\label{relxill}Comparison of best-fit values for the parameters of the models that includes different reflection component.}
 
 \begin{threeparttable}
\centering
  \resizebox{0.7\textwidth}{!}
  {\begin{minipage}{\textwidth}

\begin{tabular}{llcccc}

\hline
Model & Component &Model 4\tnote{$\dagger$}& Model 4A\tnote{$\star$} & Model 4B\tnote{$\ddag$} & Model 5\tnote{$\bullet$}\\
\hline 
{\sc edge} & E(keV) & $9.54 \pm 0.06$ &$9.56^{+0.09}_{-0.08}$ & $9.50 \pm 0.07$ & 9.57$_{-0.08}^{+0.09}$ \\
& $\tau$ & $0.010 \pm 0.009$ &0.08  $\pm$ 0.01 & 0.9 $\pm$ 0.01& 0.08 $\pm 0.01$ \\
{\sc tbabs} & nH(10$^{22}$) & $0.112 \pm 0.006$& $0.11 \pm 0.01$ & $0.13 \pm 0.02$& 0.12\tnote{*}\\ \\

{\sc bbody} & kT(keV)  &-& $0.203^{+0.008}_{-0.006}$ & $0.208^{+0.007}_{-0.01}$ & $0.19 \pm 0.003$\\
 & N(10$^{-4}$)  &-& $3.2 \pm 0.2$ & $3.0^{+0.3}_{-0.2}$& $3.20 \pm 0.06$ \\ \\
 
{\sc expabs} & E$_{Cut}$(keV) &-&$2.08^{+0.08}_{-0.15}$&-&- \\ 
{\sc relxillCp} & Index  & $0.9^{+0.3}_{-0.4}$ &  $1.38 \pm 0.5$ & $1.09^{+0.3}_{-0.4}$&- \\
{\sc xmm} & Incl(deg) & 82.5\tnote{*}& 82.5\tnote{*}& 82.5\tnote{*}&- \\
 & R$_{in}$ & 75\tnote{*}  & 75\tnote{*}  & 75\tnote{*}&- \\
% & Rout(10$^{3}$) & 1.0 \tnote{*} \\
 %& z & 0 \tnote{*} \\
 & $\Gamma$& $< 1.2$  & $1.418 \pm 0.004$ & -& -\\
 & log$_{xi}$ & $1.05 \pm 0.03$ & $1.08 \pm 0.08 $&$1.06^{+0.06}_{-0.04}$& -\\
% & Afe & 1 \tnote{*} \\
 & kTe(keV) & $2.76^{+0.02}_{-0.01}$ & $3.4 \pm 0.1 $ & -& -\\
 & refl$_{frac}$ & $1.6^{+0.1}_{-0.2}$ & $2.8 \pm 0.3$ &-& -\\
 & N(10$^{-3}$)& $1.090^{+0.009}_{-0.005}$  & $1.87 \pm 0.08$& $1.0^{+0.3}_{-0.4}$ &-\\
\\ 
{\sc expabs} & E$_{Cut}$(keV)&- & 3.0 $\pm 0.1$&-&-\\
{\sc relxillCp}& $\Gamma$  & < 2.6  &1.5$^{+0.01}_{-0.03}$ &-&-\\
{\sc NuSTAR} & log$_{xi}$  & $2.08^{+0.07}_{-0.06}$ & 2.70$^{+0.04}_{-0.03}$ & 2.37 $\pm 0.06$&-\\
 & kTe(keV)  & $4.15^{+0.04}_{-0.07}$ & 4.17$ \pm 0.07$&-&- \\
 & refl$_{frac}$ & $3.2 \pm 0.3$ &3.5$^{+0.3}_{-0.2}$ &-1\tnote{*}&-\\
 & N(10$^{-3}$)  & $1.65^{+0.02}_{-0.04}$  & 2.43$^{+0.04}_{-0.07}$ & 5.1 $\pm 0.5$&-\\ \\
 
{\sc diskline} & E(keV)&- &-&-& $6.61 \pm 0.04$   \\
{\sc xmm} & Index &- &-&-& $-2.3^{+0.3}_{-0.4}$ \\
 & R$_{in}$ &- &-&-& 75\tnote{*}   \\
 & Incl(deg) &- &-&-& $<62$   \\
 & N (10$^{-4}$) &- &-&-& $5.0 \pm 0.5$  \\
 {\sc diskline} & E(keV) &- &-&-& $6.52 \pm 0.05$   \\
 {\sc nustar} & Index &- &-&-& $ -3.0 \pm 0.8$  \\
 & N (10$^{-4}$) &- &-&-& $5.2 \pm 0.7$  \\ \\
 
{\sc pexrav} & rel$_{refl}$ &-& -&-& -1\tnote{*} \\
% & Fe$_{abund}$ & 1 \tnote{*} \\
 & N$_{xmm}$ &- &-& -& $0.0547^{+0.0078}_{-0.0088}$ \\
 & N$_{nustar}$ &- &-& -& $0.24 \pm 0.02$  \\ \\

 {\sc nthComp} & $\Gamma$ &-&-& $1.248^{+0.009}_{-0.008}$ & $1.297^{+0.003}_{-0.002}$ \\
 {\sc xmm} & kT$_{e}$(keV) &-&-&$2.98 \pm 0.04$ & $3.17 \pm 0.03$ \\
 & kT$_{bb}$(keV) &-&-& $0.65 \pm 0.04$& $0.751^{+0.005}_{-0.01}$ \\
 & N ($10^{-2}$) &-&-& $1.09^{+0.04}_{-0.03}$ & $1.03 \pm 0.01$ \\
 {\sc nthComp} & $\Gamma$ &-&-& $1.38^{+0.02}_{-0.04}$ & $1.560 \pm 0.005$\\
 {\sc nustar} & kT$_{e}$(keV) &-&-&$4.25^{+0.07}_{-0.12}$ & $ 4.97 \pm 0.02$ \\
 & kT$_{bb}$(keV) &-&-& $1.1 \pm 0.1$& $0.825^{+0.005}_{-0.042}$  \\
 & N ($10^{-3}$) &-&-& $7.1 \pm 0.4$ & $ 1.5984^{+0.0011}_{-0.0004}$ \\
 \hline 
 \\
  {\sc gaussian} & LineE(keV) & $6.513 \pm 0.007$ & $6.512 \pm 0.007$& $6.509 \pm 0.007$& $6.500 \pm 0.005$\\
{\sc XMM}& Sigma(keV) & 0.075\tnote{*}& 0.075\tnote{*}& 0.075\tnote{*}& 1.8$\times 10^{-17}$\tnote{*} \\
 & N(10$^{-4}$) & $4.6 \pm 0.2$   & $4.6 \pm 0.2$ & $5.1 \pm 0.2$& $3.0^{+0.2}_{-0.3}$ \\
 & LineE(keV) & $7.04 \pm 0.02$& $7.04 \pm 0.2$ & $7.02 \pm 0.02$& $7.12^{+0.01}_{-0.02}$\\
 & Sigma(keV) & 0.075\tnote{*}& 0.075\tnote{*}& 0.075\tnote{*}& 1.8\tnote{*}\\
 & N(10$^{-4}$) & $2.5 \pm 0.2$& $2.8 \pm 0.2$& $2.73 \pm 0.02$& $1.6 \pm 0.3 $\\ \\
 
 {\sc gaussian} & LineE(keV)  & $6.33 \pm 0.02$& 6.32 $\pm 0.03$ & 6.35 $\pm 0.02$& $6.37^{+0.02}_{-0.04}$ \\
{\sc NuSTAR}& Sigma(keV) & 0.075\tnote{*}& 0.075\tnote{*}& 0.075\tnote{*} & 1.8\tnote{*} \\
 & N(10$^{-4}$) & $5.3 \pm 0.4$ & 4.04 $\pm 0.04$ & 4.46 $\pm 0.04$& $2.5^{+0.5}_{-0.6}$\\
 & LineE(keV)& $6.81 \pm 0.04$ & 7.04\tnote{*}& 7.02\tnote{*}& $7.12$\tnote{*} \\
 & Sigma(keV)& 0.075\tnote{*}& 0.075\tnote{*}& 0.075\tnote{*}& 1.8\tnote{*}\\
 & N(10$^{-4}$) & $2.5 \pm 0.4$ &0.6 $\pm 0.3$ &1.4$^{+0.3}_{-0.4}$  & $<0.3$ \\ \\
\hline
 & $\chi^2/dof$  & 2193.0/1558 (XMM)& 3830.3/2935 & 3472.6/2932 & 3512.5/2932  \\
  &              & 1636.3/1376 (NuSTAR) & & &\\
\end{tabular}
\end{minipage}}

\begin{tablenotes}
\footnotesize
\item[*] Kept frozen during the fit.
\item[$\dagger$] \textsc{Model 4}:  \textsc{constant*edge* tbabs*(6gaussian + relxillCp)}
\item[$\star$] \textsc{Model 4A} has the same components of the \textsc{Model 4} with the addition of \textsc{bbody} and \textsc{expabs}.
\item[$\ddag$] \textsc{Model 4B} is a variant of \textsc{Model 4A} with the omission of the component \textsc{expabs} and the \textsc{Relxillcp} component used to model the reflection component only and the illuminating continuum described by \textsc{nthComp}.
\item[$\bullet$] \textsc{Model 5}: \textsc{constant*edge* TBabs*(6gaussian+ bbody + diskline + rdblur* pexrav + nthComp)}.

\end{tablenotes}
\end{threeparttable}
\end{table*}

\begin{figure}
    \centering
    \includegraphics[width=.5\textwidth]{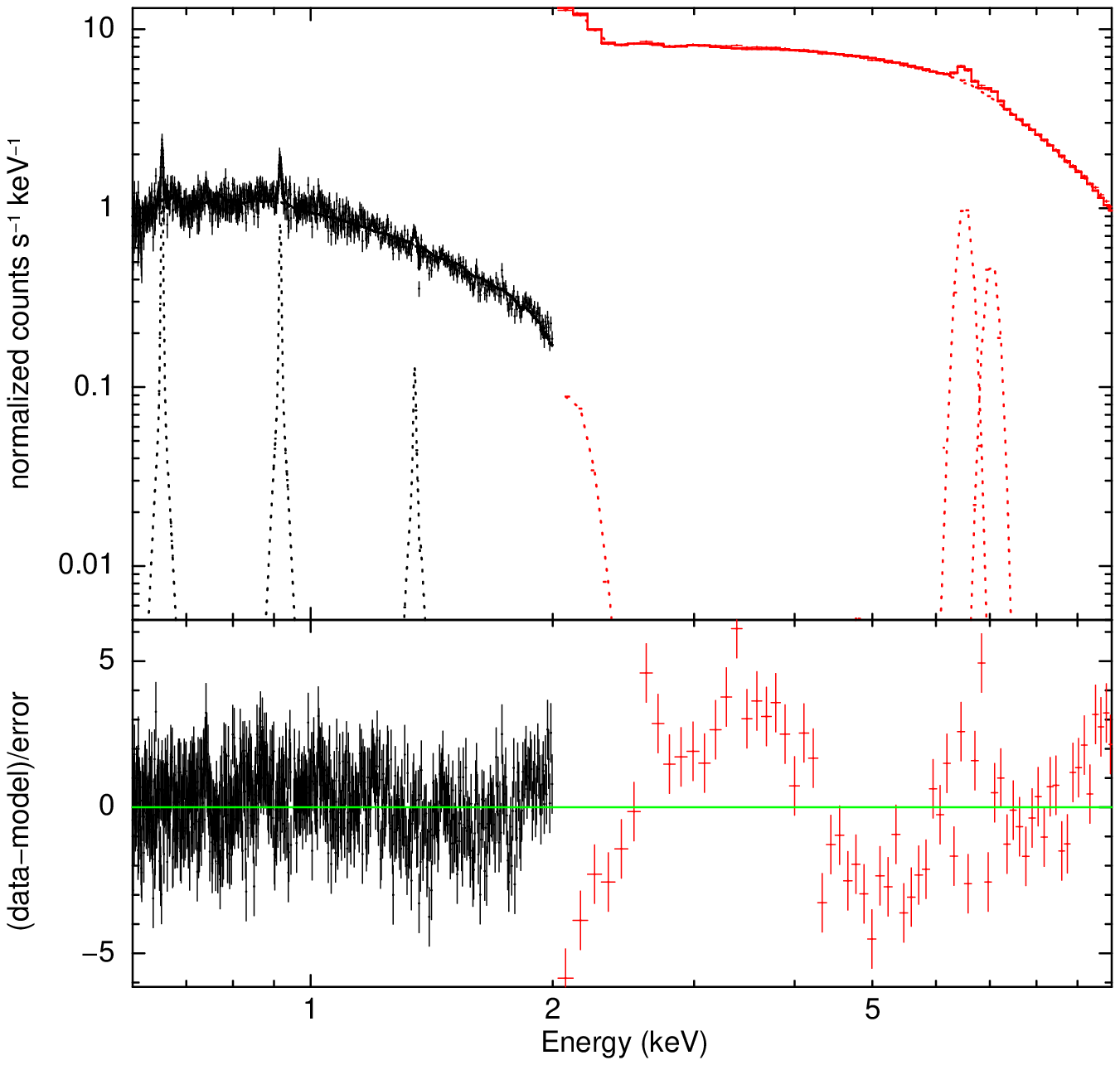}
\includegraphics[width=.5\textwidth]{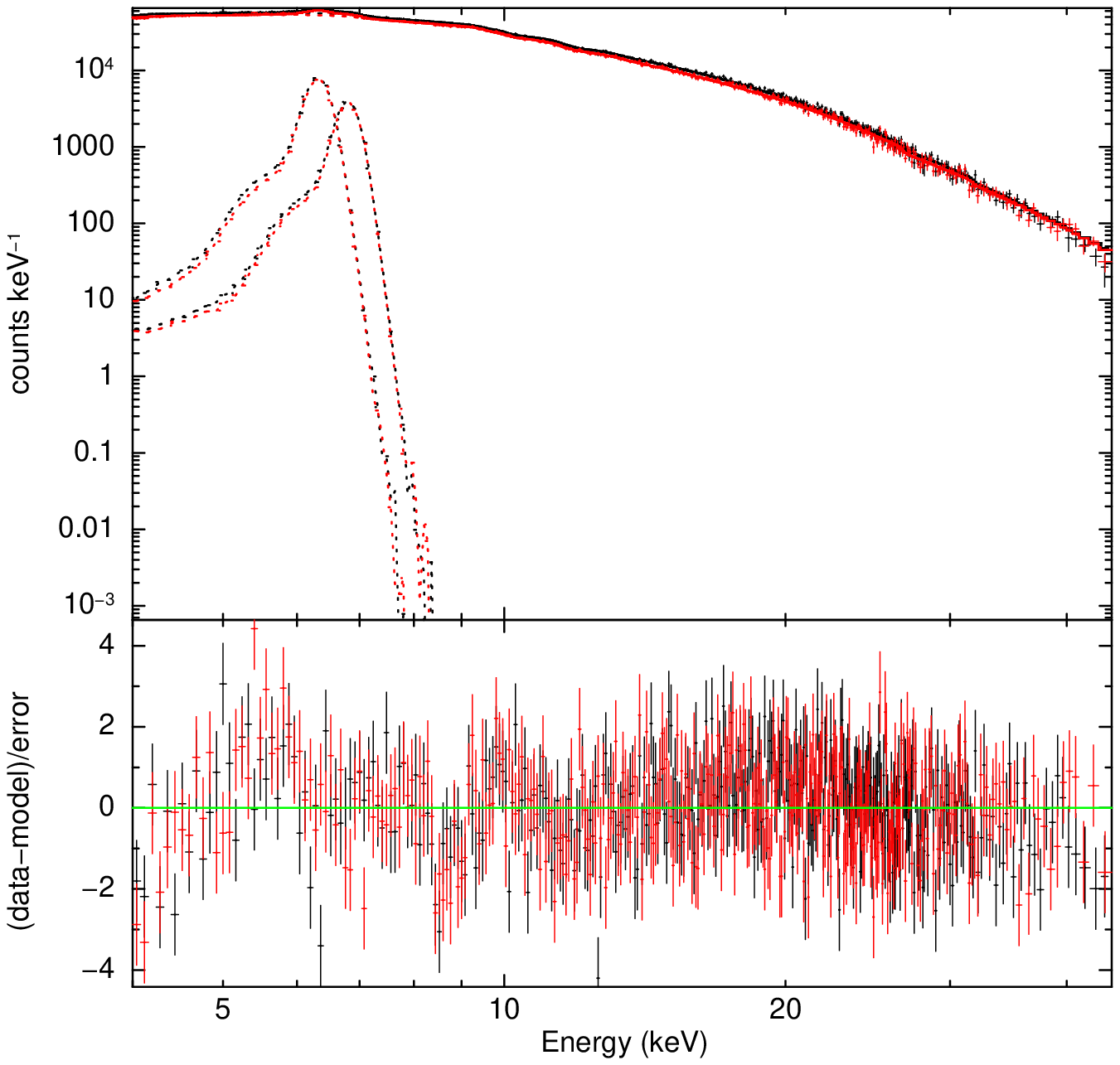}
    \caption{Spectra and residuals in units of sigma with respect to \textsc{Model 4}, defined as $constant\cdot edge\cdot tbabs\cdot(6gaussian + relxillCp)$, for the \textit{XMM-Newton} spectrum (top panel) and \textit{NuSTAR} spectrum (bottom panel), respectively. Data were rebinned for visual purposes.}
    \label{fig:Relxillcp}
\end{figure}

As discussed in Section \ref{reflection}, we tried to fit the data using two other reflection models, \textsc{RelxillCp} and a variant of Model 2, where the \textsc{pexriv} component is replaced by \textsc{pexrav}. We initially fitted the \textit{XMM-Newton} and \textit{NuSTAR} data separately using the model \textsc{RelxillCp,}  which is expected to predict the soft excess in the data that we previously described with the \textsc{bbody} component. However, it proved unable to model the spectrum of this source adequately, and as shown in Fig. \ref{fig:Relxillcp}, \textit{XMM-Newton} data show peculiar residuals in the \textit{Epic-pn} band (the $\chi^{2}$/dof is 2193.0/1558 and 1636.3/1376 for the \textit{XMM-Netwon} and \textit{NuSTAR} data, respectively).
We have to take into account that \textsc{RellxilCp} includes a Comptonisation spectrum, with a temperature of the soft seed-photons fixed at 0.06 keV, which differs from the seed-photon temperature that we find for this source (0.6-0.8 keV). To solve this problem, we added a low-energy exponential roll-off to the model using the \textsc{expabs} component in order to mimic a cut-off at the seed-photon temperature. We also tried to model our data using \textsc{RelxillCp} to describe the reflection component alone. We set the reflection fraction to -1, but allowed the normalisation to vary and linked the continuum parameters to those of the continuum fitted by the \textsc{nthcomp} component. Differently from our best-fit model, \textsc{RelxillCp} fails to reach stable values of R$_{in}$ and inclination angle. We hence fixed these values at the expected values, while the other parameters such as the ionisation parameter assume different values depending on the dataset (see Tab \ref{relxill}). We cannot use the F-test statistics to compare this fit with the fit that includes \textsc{rfxconv} because in this case, the lower chi square has the higher number of degrees of freedom.
Therefore it is preferable to choose the model with \textsc{rfxconv} because the associated fit presents a lower chi square despite the higher number of degrees of freedom.  It is clear that using the \textsc{RelxillCp} component does not ensure a better fit of the model to the data in this case. Furthermore, using \textsc{RelxillCp} alone does not fit the soft excess in the data that is modelled with the black-body spectrum, and the data still require a low-energy component to fit soft residuals in the \textit{Epic-pn} band (Fig. \ref{fig:Relxillcp}).

Although \textsc{Model 2} ensures the best fit to data with the lowest $\chi^{2}$, the component \textsc{pexriv} may  be not appropriate to model the continuum because it may not be suitable for describing a highly ionised reflection spectrum \citep{Ross_1999}. Therefore we tried to replace it with \textsc{pexrav}, an exponentially cut-off power-law spectrum reflected from neutral material \citep{1995Zidriaski}.
%We tried to replace the \textsc{pexriv} component contained in Model 2 with \textsc{pexrav}, an exponentially cut off power law spectrum reflected from neutral material \citep{1995Zidriaski} that, differently from \textsc{pexriv}, does not give the possibility to set the ionization parameter of the matter in the disc, since the latter may not be suitable for describing a ionized reflection spectrum. 
We performed the fit and kept the inner disc radius fixed at 75 R$_{g}$ because if this parameter is left free to vary, the fit is unstable and the best-fit values obtained are physically inconsistent. As shown in Tab. \ref{relxill}, we obtain reasonable values of the parameters even if the inclination angle is too small with respect to the expected value. Although the model that includes \textsc{pexrav} shows a slight improvement of the fit, by using \textsc{pexriv,} we can constrain the ionisation parameter, and the whole set of best-fit values obtained with this model is coherent with what we expect from the source. This shows that \textsc{pexriv} does not affect the results too much, also considering that the ionisation parameter obtained using the other models is quite low and \textsc{pexriv} should be able to adequately describe the effects of a mild ionisation.

\end{appendix}
\end{document}